\documentclass[universe,review,accept,moreauthors]{Definitions/mdpi} 

\firstpage{1} 
\pubvolume{1}
\issuenum{1}
\articlenumber{0}
\pubyear{2026}
\copyrightyear{2026}
\externaleditor{Firstname Lastname} 
\datereceived{21 September 2026} 
\daterevised{23 October 2026} 
\dateaccepted{31 October 2026} 
\datepublished{ } 

\Title{A Brief Review on the Statistical Properties of Fast Radio Bursts}
\Author{Yu Sang 
 $^{1}$ and Hai-Nan Lin $^{2,}$*}
\AuthorNames{Yu Sang and Hai-Nan Lin}
\address{%
$^{1}$ \quad Center for Gravitation and Cosmology, College of Physical Science and Technology, Yangzhou University, Yangzhou 225009, China; sangyu@yzu.edu.cn 
\\
$^{2}$ \quad Department of Physics, Chongqing University, Chongqing 401331, China 
}
\corres{Correspondence: linhn@cqu.edu.cn}

\abstract{Fast radio bursts (FRBs) are millisecond-duration radio transients of extragalactic origin, whose emission mechanisms and progenitors remain unresolved. Statistical analyses of the rapidly growing samples provide a promising means to constrain the underlying physics. In this review, we synthesize the statistical properties derived from current FRB catalogs, encompassing constraints on dispersion measure, redshift, energy/luminosity functions, repetition rate, and host-galaxy demographics. For repeating FRBs, we examine the statistics of energy, fluence/flux, waiting time, and discuss time-series analyses that reveal memory effects and chaotic dynamics. We further discuss the observational discovery and theoretical interpretation of periodic activity, and survey polarization properties that probe the magneto-ionic environments of FRB sources. Throughout, we emphasize robust empirical trends, account for observational biases and selection effects, and outline outstanding open questions regarding the physical origin of these enigmatic bursts.
}
\keyword{fast radio bursts; statistical properties}

\begin{document}




\section{Introduction}\label{sec:introduction}

Fast radio bursts (FRBs) are millisecond-duration, exceptionally luminous radio pulses that predominantly originate from cosmological distances, as~first established by the discovery of the ``Lorimer burst'' FRB 010724~\cite{Lorimer:2007qn} and later reinforced by subsequent bursts~\cite{Thornton:2013iua}. Over~the past decade, the~field has been  revolutionized by the advent of wide-field survey facilities such as ASKAP~\cite{Shannon:2018} and CHIME~\cite{CHIMEFRB:2021srp,TheCHIMEFRB:2026nji}, which have expanded the known source population from a handful to several thousand. These observations have unveiled a remarkable diversity in FRB properties, encompassing both apparently non-repeating events and highly active repeaters~\cite{Spitler:2016dmz,Li:2021hpl}. In~addition, they have enabled precise localizations to host galaxies~\cite{Chatterjee:2017dqg,Tendulkar:2017vuq} and have led to the definitive association of at least one FRB with a Galactic magnetar~\cite{Andersen:2020abu,Bochenek:2020zxn,Weltman:2020}. Notwithstanding this substantial progress, the~physical origin of FRBs remains among the most outstanding open questions in modern~astrophysics.

The accumulation of large, homogeneous samples has made it possible to study the statistical properties of FRBs with unprecedented precision, providing essential clues about their underlying engines and emission mechanisms. Statistical analyses of energy/luminosity functions, waiting-time distributions, and~redshift distributions have yielded robust constraints that are now central to discriminating between competing progenitor models. For~instance, the~isotropic energy distribution of FRBs has been consistently found to follow a power law with a universal index $\alpha$ $\sim$ 1.8\text{--
}1.9 across different samples~\cite{Luo:2018tiy,Lu:2019pdn,Zhang:2020ass}, while the waiting-time distributions of active repeaters exhibit bimodal features indicative of both millisecond-scale substructure and longer-term activity cycles~\cite{Li:2021hpl,Xu:2021qdn,Jahns:2022evs,Aggarwal:2021quq}. Furthermore, the~redshift distribution inferred from the CHIME catalog does not simply trace the cosmic star formation history but rather shows a significant time delay, suggesting that a substantial fraction of FRBs originate from older stellar populations or delayed formation channels~\cite{Zhang:2021kdu,Qiang:2021ljr,Lin:2024dha}. {Beyond their statistical characterization, FRBs have also emerged as independent cosmological probes, with~the dispersion measure–redshift relation providing constraints on the cosmic expansion history as well as on cosmological models and theories of gravity~\cite{Lin:2021syj,Wu:2021jyk,Zhang:2024bar,James:2022dcx,Zhuge:2025urk}.}

Several reviews have already covered the broader FRB landscape. Cordes \& Chatterjee~\cite{Cordes:2019cmq} provided a foundational overview of FRB phenomenology, propagation effects, and~source models. Petroff~et~al.~\cite{Petroff:2019tty} offered an accessible introduction to observational techniques and population statistics, and~later updated it with a comprehensive review of the major developments in following years~\cite{Petroff:2021wug}. Platts~et~al.~\cite{Platts:2018hiy} compiled a living catalog of progenitor theories, which is continuously updated to incorporate new proposals and encompasses a wide range of possible astrophysical origins. Xiao~et~al.~\cite{Xiao:2021omr} along with Zhang~\cite{Zhang:2022uzl} gave detailed accounts of the underlying physics, including radiation mechanisms and magnetar models. {Wu~et~al.~\cite{Wu:2024iyu} reviewed both statistical properties and cosmological applications of FRBs, while Wang~et~al.~\cite{Wang:2026tgi} systematically reviewed the application of FRBs in constraining the Hubble constant and dark energy.} Nevertheless, a~dedicated synthesis focused purely on the statistical characterization of FRBs remains timely, as~the rapid growth of data has solidified several key empirical results and sharpened the open~questions. 

In this review, we provide a comprehensive and critical overview of the statistical properties of FRBs, drawing on the latest observational samples. Specifically, we discuss the energy/luminosity functions of FRB sources, the~waiting‑time distributions of repeating FRBs which probe burst clustering, and~the redshift distributions that constrain the cosmic evolution of the population. In~addition, we examine the periodicity exhibited by some FRBs, which provides deep insight into the underlying physical mechanisms, as~well as the polarization statistics, which offer unique constraints on the magneto‑ionic environments of the burst sources. Throughout, we highlight the robust consensus findings, the~points of ongoing debate, and~their implications for theoretical models. By~restricting ourselves to purely statistical analyses, we aim to provide a clear, self‑contained reference for researchers seeking to understand what the ensemble properties of FRBs reveal about their physical~nature.

The statistical study of FRBs relies on samples collected by multiple telescopes across different frequency bands. Key surveys include the following. The~Parkes telescope at \mbox{$\sim$1.4 GHz} enabled the discovery of the first FRB~\cite{Lorimer:2007qn}, which established the field, and~subsequent events~\cite{Keane:2012yh,Thornton:2013iua} that confirmed their extragalactic origin. ASKAP at $\sim$1.3 GHz has localized many FRBs with sub-arcsecond precision~\cite{Bannister:2019iju,Shannon:2018}, enabling unambiguous host-galaxy associations. CHIME in the $400\text{--}800$ MHz band has released the largest FRB samples to date~\cite{CHIMEFRB:2021srp,TheCHIMEFRB:2026nji}, massively expanding the known population. UTMOST at $820\text{--}851.2$ MHz enables real-time FRB detection and rapid localization; its observing band fills the frequency gap between ASKAP and CHIME, providing complementary spectral information for understanding FRB emission physics~\cite{Farah:2019ahy,Gupta:2020uoa}. DSA-110 at $\sim$1.4 GHz delivers the largest and most uniform sample of localized FRBs to date, with~arcsecond-level precision enabling robust host-galaxy associations~\cite{Law:2023ibd} and precise measurements of FRB polarization properties~\cite{Sherman:2024iev}. FAST across $0.3\text{--}3.0$ GHz contributes high sensitivity and broad bandwidth, ideal for detailed follow-up and faint burst detection~\cite{Li:2021hpl,Uno:2025tds}. In~addition, other facilities, such as MeerKAT~\cite{Caleb:2023atr}, Apertif~\cite{Pastor-Marazuela:2024gov}, Effelsberg~\cite{Limaye:2025ghu}, and~the Green Bank Telescope (GBT)~\cite{Parent:2020hpz}, have also made significant contributions to FRB~research.

The remainder of this review is structured as follows. Section~\ref{sec:nonrepeater} summarizes the statistical properties of non-repeating FRBs, covering dispersion measure (DM), redshift, energy/luminosity functions, repetition-rate constraints, and~host-galaxy demographics. Section~\ref{sec:repeater} addresses repeating FRBs, with~emphasis on energy, fluence/flux, and~waiting-time distributions, and~also includes a time-series analysis that probes possible memory and chaotic dynamics in the central engine. Section~\ref{sec:periodicity} reviews the observational discoveries of periodic FRBs, along with detection methods and theoretical interpretations. Section~\ref{sec:polarization} discusses polarization properties, their physical interpretation, and~the implications for progenitor scenarios. Finally, Section~\ref{sec:summary} presents concluding remarks and future~prospects.

\section{Non-Repeating~FRBs}\label{sec:nonrepeater}
\unskip

\subsection{Dispersion~Measures}

The dispersion measure (DM), defined as the integral of the free-electron density along the line of sight, represents the total column density of ionized gas along the propagation path and causes the characteristic frequency-dependent delay of the received signal. The~observed DM of an extragalactic FRB is usually decomposed into four components~\cite{Deng:2013aga, Gao:2014iva, Macquart:2020lln}:
\begin{equation}\label{eq:DM_obs}
    \mathrm{DM}_{\mathrm{obs}} = \mathrm{DM}_{\mathrm{MW,ISM}} + \mathrm{DM}_{\mathrm{MW,halo}} + \mathrm{DM}_{\mathrm{IGM}} + \frac{\mathrm{DM}_{\mathrm{host}}}{1+z},
\end{equation}
where the right-hand-side terms represent the DM contributions from the Milky Way interstellar medium, Milky Way halo, intergalactic medium and host galaxy, respectively. The~$\mathrm{DM}_{\mathrm{MW,ISM}}$ term can be estimated using Galactic electron density models (e.g., NE2001 model~\cite{Cordes:2002wz,Ocker:2024rmw}, YMW16 model~\cite{Yao:2017kcp} and NE2025 model~\cite{Ocker:2026mta}). The~$\mathrm{DM}_{\mathrm{MW,halo}}$ term is less certain and is often fixed to 30$\sim$50 $\mathrm{pc\ cm}^{-3}$ \cite{Connor:2024mjg,Macquart:2020lln}, or~treated with a Gaussian prior centering at 30$\sim$80 $\mathrm{pc\ cm}^{-3}$ \cite{Prochaska:2019,Yang:2022ftm,Sang:2025mti}. Given a cosmological model, the~average contribution of the intergalactic component $\mathrm{DM}_{\mathrm{IGM}}$ can be estimated~\cite{Deng:2013aga}. However, due to large-scale structure, it is often assumed to follow a quasi-Gaussian distribution~\cite{Macquart:2020lln, Zhang:2020xoc}. Finally, the~host contribution ${\rm DM_{host}}$ is poorly known and can vary significantly from tens to 
 $\sim$1000~$\mathrm{pc\ cm}^{-3}$ \cite{Niu:2021bnl,KochOcker:2022ook}, so it is typically modeled as a lognormal distribution to allow for large variation~\cite{Macquart:2020lln,Zhang:2020mgq}.

Among the four terms, ${\rm DM_{host}}$ is the most uncertain. However, if~the other three are known, ${\rm DM_{host}}$ can be estimated by inverting Equation~(\ref{eq:DM_obs}). Using a sample of 117 well-localized FRBs, Sang \& Lin~\cite{Sang:2025mti} performed a Bayesian framework to calculate the distribution of ${\rm DM_{host}}$. They found that the natural logarithm of ${\rm DM_{host}}$ follows a normal distribution with mean $\mu_{\mathrm{host}} = 5.03\pm0.02$ and standard deviation $\sigma_{\mathrm{host}} = 0.96\pm0.03$, corresponding to a median value of $153\pm3\ \mathrm{pc\ cm}^{-3}$. A~Kolmogorov--Smirnov test yields a high \emph{p}-value (0.92), {indicating that the ${\rm DM_{host}}$ distribution is consistent with a lognormal shape.} This result provides the first independent observational validation of cosmological simulations that predict a lognormal ${\rm DM_{host}}$ distribution~\cite{Macquart:2020lln,Zhang:2020mgq}.

Given the critical role of the host-galaxy dispersion measure, a~more sophisticated model of ${\rm DM_{host}}$ is essential for FRB cosmology, since even modest uncertainties in its value can cause significant biases in determinations of cosmological parameters such as $H_0$ and $\Omega_b$. Consequently, numerous studies have explored possible correlations between ${\rm DM_{host}}$ and various host-galaxy properties, such as redshift, star formation rate (SFR), and~metallicity, etc.~\cite{Lin:2022afm,Bernales-Cortes:2025,Mo:2022qxz,Sang:2025mti,Li:2025qvl}. However, a~definitive conclusion remains~elusive.

Lin~et~al.~\cite{Lin:2022afm} analyzed a sample of 17 well-localized FRBs to search for correlations between ${\rm DM_{host}}$ and six host properties: redshift, stellar mass, SFR, galaxy age, offset, and~half-light radius. They derived ${\rm DM_{host}}$ by directly subtracting the Milky Way ISM contribution and the average IGM contribution (assuming a standard $\Lambda$CDM cosmology) from the total DM. Their analysis revealed no strong correlation with any of the examined parameters, although~a moderate positive trend with stellar mass was weakly~suggested. 
 
Bernales-Cortes~et~al.~\cite{Bernales-Cortes:2025} presented a direct empirical estimation of ${\rm DM_{host}}$ for 12 well-localized FRBs. Two methods were used to calculate ${\rm DM_{host}}$ value: one is direct estimation based on the VLT/MUSE observations of the FRB hosts, and~the other is indirect estimation based on the Macquart relation~\cite{Macquart:2020lln}. They find an average $\langle \mathrm{DM}_{\mathrm{host}}\rangle = 80\pm 11\,\mathrm{pc\,cm}^{-3}$ with a systematic uncertainty of $\sim$30\%. Positive correlations are reported between ${\rm DM_{host}}$ and both stellar mass and SFR, while no strong correlation is found with redshift or projected offset. The~study also notes a lack of correlation between the direct estimates and those inferred from the Macquart relation, suggesting missing contributions from local environments or intervening large-scale~structures.

Mo~et~al.~\cite{Mo:2022qxz} employ cosmological hydrodynamic simulations from the Illustris and IllustrisTNG projects to model the dispersion measure contributed by the interstellar and circumgalactic media of FRB host galaxies. They consider two progenitor populations: one tracing SFR (young progenitors) and the other tracing stellar mass (old progenitors). At~$z=0$, the~SFR-tracing model yields a median ${\rm DM_{host}}$ of $179\ \mathrm{pc\ cm}^{-3}$ in TNG100-1, while the stellar-mass-tracing model gives $63\ \mathrm{pc\ cm}^{-3}$. Their analysis reveals that the ${\rm DM_{host}}$ distribution deviates from a lognormal form, and~that ${\rm DM_{host}}$ increases with host stellar mass (for $M_*<10^{10.5}M_\odot$) and with redshift, while the difference between the two progenitor models diminishes at higher redshifts. In~contrast, a~similar simulation by~\cite{Zhang:2020mgq} found that ${\rm DM_{host}}$ is well consistent with a log-normal distribution and exhibits a positive correlation with~redshift.

Sang \& Lin~\cite{Sang:2025mti} also investigated correlations between ${\rm DM_{host}}$ and four host-galaxy parameters: redshift, stellar mass, SFR, and~galaxy age. A~moderate positive correlation with redshift is observed (Pearson $r$ = +0.52, \emph{p}-value $1.73\times10^{-9}$), while no statistically significant correlations are found with stellar mass, SFR, or~galaxy age. Bootstrap resampling confirms the robustness of the redshift correlation. This suggests that variations in ${\rm DM_{host}}$ are not primarily governed by global galaxy properties, but~rather by evolutionary effects (e.g., changes in metallicity or local environment density over cosmic time). The~lack of correlation with stellar mass or SFR implies that the immediate environment surrounding the FRB progenitor, such as a supernova remnant or a dense stellar cluster, may play a decisive role. One possible explanation for the positive ${\rm DM_{host}}$-redshift correlation is that an FRB at high redshift is more likely to intersect galaxy halos along the line of sight. These halos can contribute a non-negligible DM that is combined into the ${\rm DM_{host}}$ term in the calculation. In~other words, the~inferred ${\rm DM_{host}}$ value is more likely to be overestimated at higher redshifts due to the omission of the galaxy halo term along the line of sight. A~more sophisticated model of DM components should take this contribution into account~\cite{Connor:2024mjg}.

Li~et~al.~\cite{Li:2025qvl} performed spectral energy distribution (SED) and S\'ersic profile fitting for 117 FRB hosts to derive galaxy parameters. They examined correlations between the extragalactic DM (${\rm DM_E}={\rm DM_{IGM}}+{\rm DM_{host}}/(1+z)$) and host properties, finding a tight correlation between specific star formation rate (sSFR) and ${\rm DM_E}$. Although~the samples used by Sang \& Lin~\cite{Sang:2025mti} and Li~et~al.~\cite{Li:2025qvl} have large overlap, their conclusions differ significantly. The~main reason for the difference is that~\cite{Li:2025qvl} considered ${\rm DM_E}$, which includes the combined contribution of the IGM and the host galaxy, while~\cite{Sang:2025mti} focused on ${\rm DM_{host}}$. Therefore, the~positive correlation between ${\rm DM_E}$ and sSFR found by~\cite{Li:2025qvl} is not only due to ${\rm DM_{host}}$, but~also affected by ${\rm DM_{IGM}}$, the~latter being strongly~redshift-dependent.

In summary, the~observed DM of an extragalactic FRB is decomposed into Milky Way ISM, Milky Way halo, IGM, and~host-galaxy contributions. The~Galactic ISM term is now constrained by increasingly precise electron density models, while the halo term remains uncertain at the tens of pc cm$^{-3}$ level. The~IGM term follows the Macquart relation on average but is broadened by large-scale structure. The~host term is the least constrained, and~is usually described by a lognormal distribution with a median of $\sim$150~${\rm pc~cm}^{-3}$, although~its correlations with redshift and other host properties remain debated. Uncertainties in these components propagate directly into inferred redshifts, energies, and~cosmological parameters, making improved modeling of the halo and host contributions essential for future~progress.

\subsection{Redshift~Distributions}

For most FRBs without direct redshift measurements, the~extragalactic DM ($\mathrm{DM}_{\mathrm{E}} = \mathrm{DM}_{\mathrm{IGM}} + \mathrm{DM}_{\mathrm{host}}/(1+z)$) is used to infer redshift via the Macquart relation~\cite{Macquart:2020lln}. However, James~et~al.~\cite{James:2021jbo} found that above a certain DM, observational biases cause the Macquart relation to invert, implying that FRBs with the largest ${\rm DM_E}$ are not necessarily the most distant. Therefore, assuming a one-to-one correspondence between ${\rm DM_E}$ and redshift may lead to erroneous results. Furthermore, given the large uncertainties in the individual DM components, particularly the host DM contribution, the~redshift inferred from the Macquart relation carries substantial uncertainty, especially for high-redshift~FRBs.

Using the probability distributions of $\mathrm{DM}_{\mathrm{IGM}}$ and ${\rm DM_{host}}$, Tang~et~al.~\cite{Tang:2023qbg} reconstructed the $\mathrm{DM}_{\mathrm{E}}-z$ relation from 17 well-localized FRBs and applied it to the first CHIME/FRB catalog, yielding redshift estimates for hundreds of FRBs. It is found that the distributions of the ${\rm DM_E}$ and inferred redshift of the non-repeating CHIME/FRBs follow a cut-off power law, but~with a significant excess at the low-redshift range~\cite{Tang:2023qbg}. Furthermore, the~cumulative distributions of fluence and energy for the CHIME/FRBs do not follow a simple power law, but~can be well fitted by a bent power law. This can be naturally explained by the selection effect, that is, some fainter bursts may be missed by the~telescope.

The observed redshift distribution may differ significantly from the intrinsic one, as~it is strongly dependent on telescope sensitivity. The~intrinsic redshift distribution of FRBs is typically parameterized by the event rate density $dN/(dt dV)$, which denotes the number of FRB events per unit volume per unit time. The~most physically motivated model is based on the star formation history (SFH)~\cite{Madau:2014bja}; however, alternative models include power-law forms with enhancements or cutoffs, as~well as time-delayed models (e.g., Gaussian, log-normal, or~power-law delays) that account for the inspiral time of compact binary mergers~\cite{Zhang:2020ass}. Considerable effort has been devoted to constraining the event rate density of~FRBs.

The early work by Oppermann~et~al.~\cite{Oppermann:2016mzk} investigated whether the observed flux density distribution of fast radio bursts (FRBs) is consistent with a simple Euclidean model, in~which sources have a constant space density and which predicts a power-law index $\alpha = 3/2$ for the cumulative count distribution. By~performing a Bayesian analysis that combined the signal-to-noise ratios of detected bursts and the number counts from several surveys (totaling 15 FRBs), they constrained the power-law index to $\alpha \approx 0.8\text{--}1.77$ (95\% C.L.), in~good agreement with $\alpha = 3/2$. Consequently, they concluded that more complex models, such as those invoking cosmological evolution or non-uniform source densities, were not required by the data available at that time. However, this picture changed significantly as the number of discovered FRBs grew in subsequent~years.

Zhang~et~al.~\cite{Zhang:2020ass} attempted to constrain both the energy and redshift distributions of FRBs using the ASKAP (27 FRBs) and Parkes (27 FRBs) samples (mostly unlocalized). They tested two families of redshift models: one tracing the SFH and the other tracing compact binary mergers, with~the latter exhibiting a significant time delay relative to the SFH. Three delay-time models (Gaussian, log-normal, and~power-law) were considered. Through Monte Carlo simulations and Kolmogorov–Smirnov (KS) tests, they confirmed that the isotropic energy distribution follows a power law with index $\alpha \approx 1.8$; however, they were unable to constrain the exponential cutoff energy. Crucially, for~$\alpha = 1.8$, none of the redshift distribution models was rejected by either the ASKAP or Parkes data. Thus, with~the small samples available at that time, the~intrinsic redshift distribution of FRBs remained unconstrained, leaving both the SFH and merger models~viable.

Using 24 unlocalized plus 7 localized FRBs from ASKAP and 20 FRBs from Parkes, James~et~al.~\cite{James:2021oep} performed a sophisticated maximum-likelihood analysis that accounted for beamshapes, sensitivity thresholds, and~host-DM distributions. They modeled the FRB rate density as $dN/dtdV\propto [\mathrm{SFR}(z)]^n/(1+z)$ and found best-fitting $n = 1.67^{+0.25}_{-0.40}$ (spectral-index interpretation) or $n = 0.73^{+0.30}_{-0.30}$ (rate interpretation). In~all cases, a~non-evolving population ($n=0$) is ruled out at $\ge 98\%$ CL. They therefore concluded that the FRB population evolves with redshift consistent with, or~faster than, the~SFR. This result stands in apparent contrast to the subsequent CHIME-based studies~\cite{Zhang:2021kdu,Qiang:2021ljr,Lin:2023yec,Lin:2024dha}, highlighting how different telescopes, sample sizes, and~modeling choices (e.g., DM$-$z relation, host galaxy treatment, etc.) can lead to divergent~conclusions.

In stark contrast to the results of Zhang~et~al.~\cite{Zhang:2020ass} and James~et~al.~\cite{James:2021oep}, an~independent study based on the first CHIME/FRB catalog arrived at a markedly different conclusion~\cite{Zhang:2021kdu}. Using 536 FRBs from the CHIME catalog, Zhang \& Zhang~\cite{Zhang:2021kdu} tested four redshift distribution models: SFH, accumulated, delayed, and~hybrid models. Through Monte Carlo simulations and KS tests, they robustly rejected the pure SFH hypothesis and also ruled out the accumulated model, although~the latter performed slightly better than the former. They found that the time-delayed model describes the data much better, and~a hybrid model invoking both a dominant delayed population and a subdominant star formation population also fits the data. This was the first strong evidence from a large, uniform CHIME sample that FRBs do not simply trace the SFH of the~universe.

Qiang~et~al.~\cite{Qiang:2021ljr} analyzed the same CHIME catalog but adopted a different strategy: they studied six empirical redshift distribution models, namely the pure SFH model and five enhanced evolution models (PL, CPL, CSFH, TSE, and~TSRD). They fixed the energy distribution parameters ($\alpha=1.9$, $\log E_c=41$) and the selection-effect power-law index ($n=3$), and~then scanned the redshift model parameters to check whether the model could simultaneously pass the KS tests for fluence, energy, and~DM. It is found that the pure SFH model is rejected at very high confidence. In~contrast, several enhanced evolution models can be fully consistent with all three observables when model parameters are appropriately chosen. A~suppressed evolution relative to SFH (i.e., a~negative power-law index or a cutoff at high redshift) is common to all successful models, effectively requiring a ``delay''. This result broadly supports Zhang \& Zhang's rejection of SFH~\cite{Zhang:2021kdu}, while offering a wider range of viable empirical~alternatives.

Lin and Zou~\cite{Lin:2023yec} improve upon the work of Qiang~et~al.~\cite{Qiang:2021ljr} by adopting a more rigorous Bayesian inference framework. Instead of fixing energy and selection parameters, they treat all parameters as free and constrain them simultaneously using a joint likelihood of fluence, energy, and~redshift. They apply the same six redshift models and use the Bayesian information criterion (BIC) for model comparison, analyzing both a full sample and a gold sample with stricter quality cuts. Their results yield tighter and more stable constraints on the energy distribution ($\alpha\simeq1.8$–$1.9$, $\log E_c\simeq42$) and reveal that the selection-effect parameter $n$ depends strongly on the sample, while the pure SFH model is again decisively disfavored. Moreover, they find that multiple enhanced evolution models (e.g., CSFH, TSE, TSRD) can fit the data equally well depending on the sample, therefore conclude that it is still premature to make a definitive claim about the FRB~population. 

Lin~et~al.~\cite{Lin:2024dha} further refined the analysis by comparing three widely discussed time delay distributions (Gaussian, log-normal, and~power-law) against the SFH model, again using the first CHIME/FRB sample and Bayesian inference method. It was found that the log-normal delay model was mostly favored by BIC, with~a characteristic time delay of \mbox{$\sim$3\text{--}5 Gyr}. The~Gaussian and power-law delay models were also acceptable, though~less favored. The~SFH model was strongly rejected. This study provides a more direct, quantitative estimate of the delay scale, supporting the picture that most CHIME/FRBs originate from stellar populations that are significantly older than those traced by the cosmic SFR. It thus aligns with the delayed-population interpretation while moving beyond purely empirical descriptions. Furthermore, recent analysis of the second CHIME/FRB catalog also indicates that the redshift evolution of one‑off FRBs exhibits a time delay of approximately 1.4 Gyr relative to the SFH, further supporting the delayed origin scenario~\cite{Du:2026ylv}.

{In a complementary analysis, Champati \& Petrosian~\cite{Champati:2025abg} used 440 non-repeating FRBs from the first CHIME/FRB catalog and applied the non-parametric Efron--Petrosian and Lynden-Bell $C^{-}$ methods to correct for flux-limited truncation and to account for the uncertainty in DM-based redshifts through lower, mean, and~upper redshift samples. They found $\sim$3$\sigma$ evidence for luminosity evolution, a~broken power-law luminosity function, and~a comoving formation rate that declines rapidly with redshift and differs significantly from the SFH. The~inferred rate evolution closely resembles that of short gamma-ray bursts, whose progenitors are delayed relative to the SFH, thereby providing independent support for a delayed formation scenario and, more broadly, for~magnetar or compact-binary-merger progenitors of FRBs.}

Zhang~et~al.~\cite{Zhang:2026dqt} performed an independent population study of non‑repeating FRBs using over one thousand events from the second CHIME/FRB Catalog, employing a self‑consistent framework that combines backward non-parametric inference (weighted Lynden‑Bell $C^{-}$ estimator) and forward Monte Carlo population synthesis while carefully accounting for the fuzzy selection function and the probabilistic DM-redshift estimations. Their backward non‑parametric inference recovers an intrinsic redshift distribution peaking at $z \sim 1$, significantly below the SFH peak at $z \sim 1.7$, which corresponds to a time delay of $\sim$2 Gyr. Their forward population synthesis further confirms that the SFH model is strongly disfavored. These results independently corroborate the delayed origin scenario and strengthen the conclusion that the non‑repeating FRB population does not directly trace the cosmic~SFH.

{Although most analyses of the first CHIME/FRB catalog find that a pure SFH model is disfavored and that either suppressed evolution or a Gyr-scale delay is preferred, controversy remains. For~example, Shin~et~al.~\cite{Shin:2022crt} used the first CHIME/FRB catalog with an injection-based selection function and a hierarchical Bayesian forward model to infer the FRB energy and distance distributions. They found that, once CHIME selection effects are accounted for, the~evolution of the FRB volumetric rate relative to the SFH is not significantly constrained. This result is broadly consistent with Wang~et~al.~\cite{Wang:2026tco}, showing that claims of a Gyr-scale delay are sensitive to the treatment of selection effects.} Employing a forward-modeling hierarchical Bayesian analysis that self-consistently incorporates the CHIME injection-based selection function alongside baseband fluences and localized host redshifts, Wang~et~al.~\cite{Wang:2026tco} find that the FRB volumetric rate peaks at essentially the same redshift as the SFH, with~a mean delay of only $0.1\text{--}0.3$ Gyr, consistent with zero delay at $2\sigma$. They argue that previous claims of long delays are likely driven by simplified treatments of selection effects. This contrasting result underscores that, while a delayed origin remains the prevailing consensus, the~exact delay scale and progenitor channel are still actively debated; further progress will require larger samples of localized hosts and continued cross-validation among independent methodological~approaches.

\subsection{Energy/Luminosity~Functions}

The isotropic-equivalent energy or luminosity of FRBs is typically modeled by a power law with an exponential cutoff, also known as the Schechter function:
\begin{equation}
\frac{dN}{dE} \propto \left(\frac{E}{E_{\rm max}}\right)^{-\alpha} \exp\left(-\frac{E}{E_{\rm max}}\right).
\end{equation}
Several 
studies have been devoted to constraining the power-law index $\alpha$ and cut-off energy $E_{\rm max}$ using various samples and methods, and~the main results are summarized in Table~\ref{tab:energy_function}.

\begin{table}[H]
\caption{{Constraints 
on the FRB energy/luminosity function. Here, $\alpha$ is the power-law index of differential distribution. }}\label{tab:energy_function}
\begin{adjustwidth}{-\extralength}{0cm}
\small
\begin{tabular}{l >{\raggedright\arraybackslash}m{4.4cm} >{\raggedright\arraybackslash}m{4.1cm} >{\raggedright\arraybackslash}m{1.6cm} l}
\toprule
\textbf{Reference} & \textbf{Sample (\# of FRBs)} & \textbf{Energy/Luminosity Band} & \boldmath$\alpha$ & \boldmath$E_{\rm max}$ \textbf{or} \boldmath$L_{\rm max}$ \\ \midrule
Luo~et~al.~\cite{Luo:2018tiy} & Mixed sample (33) & $\log(L/{\rm erg~s^{-1}}) = 40\text{--}45.5$ & 1.2$\sim$1.8 & $L_{\rm max}\sim 2\times 10^{44}~ {\rm erg~s^{-1}}$ \\ \midrule
Lu \& Piro~\cite{Lu:2019pdn} & ASKAP (23) & $\log(E/{\rm erg~Hz^{-1}}) = 31\text{--}34.5$ & $1.6\pm0.3$ & $E_{\rm max}\sim 10^{34.1}~ {\rm erg~Hz^{-1}}$ \\ \midrule
Zhang~et~al.~\cite{Zhang:2020ass} & ASKAP+
Parkes (27+27) & $\log(E/{\rm erg})\sim 39\text{--}44 $ & 1.8 {(fixed)}  & $\log(E_{\rm max}/{\rm erg})\sim 41.5$ \\ \midrule
James~et~al.~\cite{James:2021oep} & ASKAP+Parkes (31+20) & $\log(E/{\rm erg})\sim 30\text{--}42 $ $^a$ & $2.09_{-0.10}^{+0.14}$ & $\log(E_{\rm max}/{\rm erg})\sim 41.7$\\ \midrule
Zhang \& Zhang~\cite{Zhang:2021kdu} & CHIME (492) & $\log(E/{\rm erg})\sim 35\text{--}42$ & 1.9 {(fixed)} & $\log(E_{\rm max}/{\rm erg})\sim 41.5$ \\ \midrule
Qiang~et~al.~\cite{Qiang:2021ljr} & CHIME (452) & $\log(E/{\rm erg})\sim 35\text{--}43$ & 1.9 {(fixed)} & $\log(E_{\rm max}/{\rm erg})\sim 41.0$ \\ \midrule
Lin \& Zou~\cite{Lin:2023yec} & CHIME (Full: 436, Gold: 236) & $\log(E/{\rm erg})\sim 37\text{--}43$ & 1.8$\sim$1.9 & $\log(E_{\rm max}/{\rm erg})\sim 42$ \\ \midrule
Lin~et~al.~\cite{Lin:2024dha} & CHIME (Full: 436, Gold: 236) & $\log(E/{\rm erg})\sim 37\text{--}43$ & 1.8$\sim$2.1 & $\log(E_{\rm max}/{\rm erg})\sim 42$ \\ 
\bottomrule
\end{tabular}

\begin{adjustwidth}{+\extralength}{0cm}
{\noindent{\small{Note: $^a$ The lower limit of $\log(E/{\rm erg})\sim 30$  is a fixed parameter in the analysis of James~et~al.~\cite{James:2021oep}.}}}
\end{adjustwidth}
\end{adjustwidth}
\end{table}
\vspace{-3pt}

Despite the diverse methodologies, samples, and~modeling assumptions, all the works converge on the following robust~conclusions:
\begin{itemize}
    \item A universal power-law index $\alpha \sim 1.8\text{--}1.9$. The~most secure result is that the FRB isotropic energy distribution follows a power-law form, $dN/dE \propto E^{-\alpha}$, with~the index consistently converging to $\alpha \approx 1.8\text{--}1.9$. This result shows remarkable stability across data samples, ranging from the 33 FRBs analyzed by Luo~et~al.~\cite{Luo:2018tiy} to the hundreds of FRBs in the CHIME catalog studied by~\cite{Lin:2023yec,Lin:2024dha}. It is independent of the assumed Galactic electron models, host DM~distributions.
    
    \item A high-energy cutoff at $\log(E_{\rm max}/{\rm erg)} \sim 41\text{--}42$. While early small-sample studies~\cite{Zhang:2020ass} could not place tight constraints on the high-end cutoff, the~advent of the large, uniform CHIME sample has firmly pinned down the cutoff energy to \mbox{$\log(E_{\rm max}/{\rm erg}) \approx 42$} \cite{Zhang:2021kdu,Qiang:2021ljr,Lin:2023yec,Lin:2024dha}, which corresponds to an isotropic luminosity of \mbox{$\sim$$10^{45}$ erg s$^{-1}$} for millisecond-duration bursts. This finding is corroborated by James~et~al.~\cite{James:2021oep} based on ASKAP and Parkes samples, who derived $\log(E_{\rm max}/{\rm erg}) = 41.70$ using a detailed maximum-likelihood approach.
\end{itemize}

In summary, the~FRB isotropic energy function is empirically well described by a power law with a high-energy cutoff at $E_{\rm max} \sim 10^{42}$ erg. This result is robust and independent of the assumed redshift distribution models. However, its physical interpretation is intimately tied to the underlying redshift evolution: the observational data strongly disfavor a young-magnetar population that traces the SFH and instead point toward an older, delayed stellar population, such as merging compact binaries or old magnetars produced via alternative formation~channels.

\subsection{Repetition Rate~Constraints}

FRBs are typically divided into repeaters and apparent non-repeaters. The~latter, however, may not be intrinsically non-repeating; their bursts might simply be undetected by telescopes or emit in the future. Deep follow-up observations provide stringent upper limits on their repetition rates, which are essential for determining whether all FRBs are inherently repeaters or whether a genuinely non-repeating population~exists.

Good~et~al.~\cite{Good:2022bze} conducted follow-up observations of {two known repeating FRBs, and~seven non-repeating FRBs with distinctive downward-drifting frequency structure often observed in repeating FRBs}, all discovered with CHIME/FRB project, using the Arecibo telescope, but~no additional bursts were detected for either source. They derived upper limits on the repetition rates for all sources, under~both Poisson and Weibull assumptions, on~the order of $\lambda = 10^{-2}$ to $10^{-1}$~h$^{-1}$. These rates are much lower than those recently published for notable repeating sources such as FRB~20121102A ($\sim$736 day$^{-1}$) \cite{Zhang:2021ztz} and FRB~20201124A ($\sim$10 day$^{-1}$) \cite{Lanman:2021yba}, suggesting the existence of a \mbox{low-repetition~subpopulation.}

Uno~et~al.~\cite{Uno:2025tds} performed follow-up observation of 36 apparent non-repeating FRBs with FAST, each for 10 min, achieving a typical $7\sigma$ fluence limit of $\sim$0.013~Jy~ms, and~no bursts with an SNR above 7 were detected. The~derived upper limits on the repetition rates are $\sim$$10^{-2.6}$ to $10^{-0.22}$~h$^{-1}$ under a Poisson assumption, and~$\sim$$10^{-2.3}$ to $10^{-0.25}$~h$^{-1}$ under a Weibull process. These limits are among the most stringent reported to date, benefiting from a sample size five times larger than that of previous studies~\cite{Good:2022bze}. If~all FRB sources are repeaters, the~majority must have repetition rates below $10^{-2}$~h$^{-1}$, which would require exceptionally long active periods to account for the observed population of apparently one-off~bursts.

Yang~et~al.~\cite{Yang:2025dpl} conducted a systematic follow-up campaign on 27 of the 81 non-repeating FRBs identified in the Parkes Transient Database, detecting no additional bursts from any of these sources. Combining these non-detections with archival observations, they derived stringent upper limits on repetition rates for bursts above 1~Jy. Under~a Poisson process, the~limits range from $\sim$$10^{-3.5}$ to $10^{-1.9}$~h$^{-1}$; under a Weibull process, they range from $\sim$$10^{-3.4}$ to $10^{-1.5}$~h$^{-1}$. These constraints are roughly an order of magnitude tighter than those reported by Uno~et~al.~\cite{Uno:2025tds}. By~employing consistent observational setups and analysis methods across all sources, the~derived limits converge to a narrow, well-defined interval, suggesting that these FRBs constitute a relatively homogeneous population with extremely low intrinsic burst~activity.

Repeating FRBs account for only a few percent of the CHIME/FRB sample, yet this figure is almost certainly a lower limit imposed by selection effects. By~correcting for these biases, Yamasaki~et~al.~\cite{Yamasaki:2023dlb} demonstrate a marked decline in the detection rate of apparently non-repeating FRBs as CHIME's operational baseline lengthens, implying that many such sources would eventually exhibit repetition given sufficient monitoring. Their simple population modeling further suggests that the true repeater fraction exceeds 50\% at the 99\% confidence level, substantially above the observed value, and~even consistent with unity. This overwhelming prevalence of repeaters has remained hidden solely because of their extremely low repetition activity. Consequently, the~inferred volume number density of repeaters may exceed that of the observed population by up to four orders of~magnitude.

Beniamini and Kumar~\cite{Beniamini:2025egd} developed a model‑independent framework demonstrating that FRB sources follow a Zipf-like distribution, in~which the source number density is approximately inversely proportional to the burst rate above a fixed energy threshold. This inverse scaling holds even though the burst rate and number density each individually span many orders of magnitude. In~addition, they find that the repeater fraction increases only mildly with improved sensitivity or longer exposure times, and~this weak dependence could be easily misinterpreted as evidence against universal repetition. Overall, their results favor the view that a single progenitor population, most likely magnetars, can naturally explain the full observed diversity of FRB activity, ranging from very inactive sources such as SGR 1935+2154 to the most prolific~repeaters.

Ai~et~al.~\cite{Ai:2020wnm} proposed a s tatistical method to investigate whether genuine non-repeaters exist. They introduced a parameter $T_c$, defined as the timescale at which the accumulated number of non-repeating sources becomes comparable to the total number of repeating sources. Through Monte Carlo simulations, they demonstrate that for a finite $T_c$, the~observed repeater fraction $F_{\mathrm{r,obs}}$ should rise with time until reaching a peak, and~subsequently decline, if~genuine non-repeaters really exist. In~contrast, $F_{\mathrm{r,obs}}$ would either increase monotonically or saturate if all sources are repeaters. Measuring this temporal evolution requires continuous monitoring of FRBs with wide-field radio telescopes, and~detection of a smaller peak value $F_{\rm r,obs,p}<0.04$ would disfavor the ansatz that ``all FRB sources repeat''. Subsequent observations from the CHIME telescope indicate that the repeater fraction tends to an equilibrium of $0.026$, suggesting the existence of genuine non-repeaters~\cite{CHIMEFRB:2023myn}.

In summary, while deep follow-up observations continue to place stringent upper limits on the repetition rates of apparently non-repeating FRBs, the~true fraction of repeating sources remains uncertain. The~emerging picture suggests that many, perhaps all, FRB sources may repeat, albeit at extremely low rates, and~that the apparent dichotomy between repeaters and non-repeaters is likely primarily a consequence of observational sensitivity and limited monitoring~time.

\subsection{Host-Galaxy~Demographics}

The identification and characterization of FRB host galaxies are crucial for understanding the progenitors and environments of these enigmatic bursts. A~study of 23 well-localized FRB hosts (6 repeaters and 17 apparently non-repeaters) found no statistically significant differences in global host properties between the two classes \citep{Gordon:2023cgw}. It is found that most FRB hosts (20 of 23) lie on the star-forming main sequence, but~at least three FRBs originate in less active environments. Although~the overall host properties of repeaters and non-repeaters are statistically indistinguishable, repeaters tend to reside in lower-mass galaxies, while non-repeaters favor more optically luminous hosts. This aligns with the finding that repeating FRBs typically occupy the faint, low-mass end of the host distribution \citep{Tendulkar:2017vuq,Niu:2021bnl,Hewitt:2024sav}. A~remarkable exception is the repeater FRB 20240209A, which resides in a massive, quiescent elliptical host galaxy~\cite{Eftekhari:2024bmi}. The~non-repeater FRB 20230708A is another exception, lying in a low stellar mass, low SFR, and~very low metallicity dwarf galaxy, making it the lowest‑luminosity non‑repeating FRB host discovered to date~\cite{Muller:2025mcf}.

Subsequent studies with larger samples have refined this picture. An~extensive analysis of the stellar mass, SFR, and~redshift distributions of 51 FRB hosts suggests that FRB progenitors are more likely to trace SFR than stellar mass \citep{Loudas:2025cnd}. Nevertheless, a~hybrid model allowing up to $\sim$50\% of FRBs to track stellar mass remains consistent with the data, hinting at the coexistence of multiple formation channels. Using the same sample but a different statistical approach, Horowicz \& Margalit~\cite{Horowicz:2025rhc} found that the hypothesis of FRBs tracing either SFR or globular cluster mass is disfavored, while a mixed model where FRBs follow a linear combination of SFR and stellar mass is favored instead. Statistical analyses of the first CHIME/FRB catalog further confirm that the FRB population does not simply trace the SFR; rather, time delays are required to reconcile the data with the observed distribution~\cite{Qiang:2021ljr,Lin:2023yec,Lin:2024dha}.

Detailed observations of nearby FRB hosts offer further insight. An~analysis of 18 local hosts (7 repeaters and 11 non‑repeaters, all at $z<0.1$) shows that all are spiral or late‑type galaxies \citep{Bhardwaj:2023vha}, implying that core‑collapse supernovae are likely the dominant formation channel in the local Universe. Furthermore, the~host properties of repeaters and apparent non‑repeaters do not differ significantly within this~sample.

The diversity of FRB host galaxies challenges the notion of a single progenitor pathway. A~striking example is repeating FRB~20240209A, which resides in a massive, quiescent elliptical galaxy with $\log(M_*/M_\odot)=11.35$ and a mass-weighted age of $\sim$11 Gyr \citep{Eftekhari:2024bmi}. This is the most massive and oldest FRB host known, and~the first confirmed elliptical host. The~presence of such a quiescent early-type galaxy within a transient class otherwise dominated by star-forming spirals echoes the host environments of short-duration gamma-ray bursts, type Ia supernovae, and~ultraluminous X-ray sources. Candidate progenitors include magnetars formed through merging binary neutron stars or white dwarfs, or~via accretion-induced collapse. Together with FRB~20200120E, localized to a globular cluster in M81~\cite{Kirsten:2021llv}, these findings provide strong evidence that a subset of FRBs may originate from processes distinct from massive-star core~collapse.

Complementary results from 30 FRB host galaxies discovered by the DSA-110 show a significant deficit of low-mass FRB hosts compared to the occurrence of star formation in the universe, implying that FRBs are a biased tracer of star formation, preferentially selecting massive star-forming galaxies \citep{Sharma:2024fsq}. This bias may be driven by galaxy metallicity, which is positively correlated with stellar mass. Metal-rich environments may favor the formation of magnetar progenitors through stellar mergers, as~higher metallicity stars are less compact and more likely to fill their Roche lobes, leading to unstable mass transfer. Merger remnants are thought to have the requisite internal magnetic-field strengths to result in magnetars, suggesting that core-collapse supernovae of merger remnants preferentially form magnetars \citep{Sharma:2024fsq}.

Studies of gas-phase metallicity in FRB hosts provide further constraints on progenitor models. Based on a large uniformly selected sample comprising 31 hosts with measured oxygen abundances, Yamasaki~et~al. \citep{Yamasaki:2025huw} found that FRB host galaxies span a wide metallicity range and are closely follows the SFR-weighted mass--metallicity relation of star-forming galaxies. Contrary to earlier claims~\cite{Sharma:2024fsq} that proposed a possible metallicity dependence in FRBs, no clear metallicity preference is found in this sample, indicating that metallicity alone does not regulate FRB production. This implies that FRBs can arise even in low-metallicity, high-redshift systems. A~marginal offset from the fundamental metallicity relation is observed, likely driven by suppressed SFRs at fixed mass and metallicity. This may reflect a post-starburst phase following galaxy interactions, in~which FRB progenitors formed during the starburst and produce FRBs after a $100\text{--}500$ Myr delay. Such a scenario favors binary evolution channels over core-collapse~supernovae.

The rapid growth of well-localized samples continues to reshape this picture. The~CHIME/FRB Collaboration~\cite{CHIMEFRB:2025ggb} localized 81 new FRBs to arcsecond precision using very long baseline interferometry (VLBI) between CHIME and the KKO outrigger station. Through probabilistic host association, the~authors securely identified 21 new FRB host galaxies and compiled spectroscopic redshifts for 19 of them, 15 of which were newly obtained. This catalog significantly increases the statistics of the Macquart relation at low redshifts ($z < 0.2$) and highlights several interesting cases, including a repeating FRB in a galaxy merger, FRBs toward galaxy clusters, and~a luminous persistent radio source candidate. These additions improve the statistical leverage of host-galaxy demographics at low redshift and provide a homogeneous reference sample for comparisons with higher-redshift~populations.

In summary, the~diversity of host environments, ranging from dwarf star-forming galaxies to massive quiescent ellipticals and globular clusters, strongly suggests multiple progenitor channels. Notably, the~detection of an FRB-like burst from the Galactic magnetar SGR~1935+
2154 confirmed that at least some FRBs originate from young magnetars formed via core-collapse supernovae \citep{Bochenek:2020zxn}. However, the~time delay of $3\text{--}5$ Gyr inferred from population studies implies that a substantial fraction of FRBs must arise from older progenitors, such as old magnetars formed via accretion-induced collapse or binary mergers \citep{Zhang:2021kdu,Lin:2024dha}. The~accumulating evidence from host-galaxy demographics points toward a complex picture where FRBs can arise from multiple formation channels, including both young and old stellar~populations.

\subsection{Summary for Non-Repeating FRBs}

Considerable advances in the study of FRBs have occurred over the past decade. These have been driven by large, homogeneous surveys, particularly those conducted with CHIME/FRB, and~by the growing sample of well‑localised bursts with host‑galaxy associations. From~this body of work, several robust empirical facts have clearly~emerged.

First, the~dispersion measure of extragalactic FRBs can be decomposed into Galactic, IGM, and~host contributions. The~host contribution, $\mathrm{DM}_{\mathrm{host}}$, is well described by a lognormal distribution with a median of $\sim$150~$\mathrm{pc\,cm^{-3}}$, in~agreement with cosmological simulations \citep{Mo:2022qxz,Sang:2025mti}. However, its correlation with host properties remains contentious: while some studies find positive trends with redshift, stellar mass or SFR \citep{Sang:2025mti,Li:2025qvl,Bernales-Cortes:2025}, others report no significant correlations \citep{Lin:2022afm}. This discrepancy likely arises from sample selection, different calculation methods of $\mathrm{DM}_{\mathrm{host}}$, and~the difficulty of separating $\mathrm{DM}_{\mathrm{host}}$ from the IGM contribution. A~definitive understanding of the physical drivers of $\mathrm{DM}_{\mathrm{host}}$, whether they are halo gas, local environment, or~galaxy‑scale properties, is still~lacking.

Second, the~redshift distribution of non-repeaters has been intensively studied using the CHIME catalog. {The majority of} analyses converge on the rejection of the pure SFH model \citep{Zhang:2021kdu, Qiang:2021ljr, Lin:2023yec, Lin:2024dha, Du:2026ylv}. Instead, the~data require either an evolution that is suppressed relative to SFH, or~a significant time delay (typically several Gyr) between star formation and FRB emission. This suggests that the dominant progenitor population of CHIME/FRBs is not young magnetars from core-collapse supernovae, but~rather older systems, such as merging compact binaries, old magnetars formed via accretion-induced collapse, or~other delayed channels. Nevertheless, the~precise functional form of the delay (Gaussian, log-normal, or~power-law) remains poorly constrained, and~different models can fit the data equally well depending on the sample and analysis method. {A minority of studies, however, argue that the apparent Gyr-scale time delay can be explained by CHIME selection effects, and~that little or no time delay is required once selection effects are properly taken into account~\cite{Shin:2022crt,Wang:2026tco}. Therefore, young magnetars from core-collapse supernovae cannot be ruled out by current data.}

Third, the~isotropic-equivalent energy function of non-repeaters is robustly described by a power law with an exponential cutoff, with~a universal index $\alpha \simeq 1.8$ to $1.9$ and a cutoff energy $\log(E_{\rm max}/{\rm erg}) \simeq 42$ \citep{Zhang:2021kdu, Qiang:2021ljr, Lin:2023yec, Lin:2024dha}. This result is remarkably stable across telescopes, samples, and~modeling assumptions. The~physical interpretation of this cutoff, however, is intertwined with the redshift evolution: for a delayed population, the~inferred volumetric rate and energy distribution may have different implications than for an SFH-tracing~one.

Fourth, deep follow-up campaigns targeting apparently non-repeating FRBs have yielded stringent upper bounds on their repetition rates, typically $\lesssim$$10^{-2}~\mathrm{h^{-1}}$ \citep{Good:2022bze, Uno:2025tds, Yang:2025dpl}. Although~these limits are considerably lower than the rates measured for known repeaters, population models indicate that the intrinsic repeater fraction could be much larger than the observed few per cent, and~might even approach unity \citep{Yamasaki:2023dlb, Beniamini:2025egd}. Hence, the~apparent dichotomy between repeaters and non-repeaters could be an observational selection effect, with~most sources being repeaters of very low activity. Nevertheless, temporal evolution analyses of the repeater fraction still allow for the existence of a genuine class of truly one-off events \citep{Ai:2020wnm, CHIMEFRB:2023myn}.

Fifth, host-galaxy demographics reveal a remarkable diversity: FRB hosts range from low-mass, star-forming dwarf galaxies to massive, quiescent ellipticals, and~even globular clusters \citep{Gordon:2023cgw, Eftekhari:2024bmi, Kirsten:2021llv}. While repeaters tend to prefer lower-mass hosts, the~overall host properties of repeaters and apparent non-repeaters are statistically indistinguishable in many samples \citep{Gordon:2023cgw, Bhardwaj:2023vha}. This diversity, together with the discovery of FRB-like bursts from the Galactic magnetar SGR~1935+2154 \citep{Bochenek:2020zxn}, points to a coexistence of multiple formation channels. These include young magnetars from core collapse, old magnetars from accretion-induced collapse or binary mergers, and~possibly other pathways.

\section{Repeating~FRBs}\label{sec:repeater}

While Section~\ref{sec:nonrepeater} focused on the ensemble properties of apparently non-repeating events, a~pivotal subset of the FRB population exhibits recurrent bursting activity from the same source. These repeating FRBs offer a unique observational advantage, e.g.,~they allow for continuous monitoring, detailed time-domain analyses, and~robust statistical modeling of individual source dynamics over long baselines. The~emission from these repeaters often exhibits complex temporal and energetic clustering that simple Poisson processes or single power-law models fail to capture. In~this section, we review the statistical characterization of repeating FRBs. We first introduce the statistical methodologies to characterize these repeating events. Subsequently, we systematically review their energy, flux density, fluence, and~waiting-time distributions. Finally, we explore nonlinear time-series analyses that probe the underlying physical mechanisms, providing essential clues to the central engines of these highly active~transients.

\subsection{Statistical Methods for Repeating~FRBs}

We begin by reviewing the statistical methodologies commonly employed to characterize the statistical properties of repeating FRBs. Traditionally, the~energy distribution of these transients has been modeled as a power law, while the waiting time between successive bursts is assumed to follow a Poisson process. However, accumulating observational evidence from active repeaters has revealed that these simple models are inadequate to capture the full complexity of their statistical behavior. More sophisticated distributions are needed to account for observed deviations, such as energy turnovers, breaks, and~temporal clustering. Below, we outline the principal models that have been successfully applied to the energy and waiting-time distributions of repeating FRBs. Since fluence and flux density scale approximately with burst energy, the~same models can be used for their~distributions.

The simplest and most widely adopted model for the energy distribution of repeating FRBs is the simple power law (SPL), expressed in differential form as $dN/dx \propto x^{-\alpha}$, or~equivalently in cumulative form as~\cite{Wang:2016lhy,Wang:2019sio,Lin:2019ldn}
\begin{equation}
N({>}x) = A (x^{-\alpha+1}-x_c^{-\alpha+1}),~~ x<x_c
\end{equation}
where $\alpha$ is the (differential) power-law index, and~$x_c$ is the cutoff value above which $N({>}x_c)=0$. This model originates from the empirical scaling laws observed in various astrophysical phenomena, such as solar flares and gamma-ray bursts, and~its appeal lies in its simplicity and the direct physical interpretation: a constant power-law slope suggests a scale-invariant emission process with no preferred energy scale. For~repeating FRBs, early studies of sources like FRB 20121102A often adopted a power-law fit to the burst energy distribution, yielding index values typically in the range $\alpha \sim 1.5\text{--}2.5$ \cite{Wang:2016lhy,Wang:2019sio,Lin:2019ldn,Zhang:2021ztz}. Despite its wide usage, the~single power law is increasingly recognized as insufficient for many repeaters, because~it fails to account for observed turnovers at low energies (due to detection thresholds) and breaks or cutoffs at high energies (due to intrinsic physical limits or sample sparsity) \cite{Lin:2019ldn,Zhang:2021ztz,Sang:2023zho,Lyu:2020hoz}. Nevertheless, it remains a benchmark model against which more complex forms are compared, and~it provides a useful first-order characterization of the energy scale and burst rate of repeating~sources.

The bent power-law (BPL) model provides a phenomenological description of distributions that deviate from a single power law at both the low and high extremes. It has a form of~\cite{Lin:2019ldn}
\begin{equation}
N({>}x) = B \left[ 1 + \left( \frac{x}{x_b} \right)^{\beta} \right]^{-1},
\end{equation}
where $x_b$ is the median value of $x$, i.e.,~the number of data points lying above $x_b$ is equal to that below $x_b$. $B$ is the total number of data points, and~$\beta$ is the power-law index in the limit $x\gg x_b$. The~BPL model effectively acts as a smoothly connected piecewise power law, featuring a flat tail at the low end (power-law index $p=0$) and a power-law behavior at the high end (power-law index $p=\beta$). The~BPL model was first proposed to describe the power density spectra of gamma-ray bursts~\cite{Guidorzi:2016ddt}, and~has since been widely applied to the statistical distributions of various transients, such as soft gamma repeaters (SGRs)~\cite{Chang:2017bnb,Sang:2021cjq} and FRBs~\cite{Lin:2019ldn,Wang:2022gmu,Sang:2023zho}, as~well as to the spectra of active galactic nuclei (AGN) \cite{Panagiotou:2022hqr,Lefkir:2025jra}. Specifically for repeating FRBs, Lin \& Sang~\cite{Lin:2019ldn} pioneered the use of the BPL model to the cumulative distributions of fluence, flux density, energy and waiting time of FRB 20121102A, finding that it yields an excellent fit across a wide dynamic~range.

In addition to the BPL model, several other functional forms have been introduced to capture specific features of the energy distribution. The~thresholded power-law (TPL) model, originally developed to describe solar and stellar flares~\cite{Aschwanden:2015}, has been shown to fit the energy distributions of both FRBs and magnetar bursts~\cite{Cheng:2019ykn,Sang:2021cjq,Sang:2023zho}. Its differential form is given by $N(x)dx = n_0 (x_0 + x)^{- \gamma } dx$ for $x_1\leq x\leq x_2$, where $x_0$ is a free parameter that controls the position of the turnover point, $n_0$ is a normalization constant, $x_1$ and $x_2$ are the lower and upper cutoffs, respectively. The~corresponding cumulative distribution is
\begin{equation}
   N({>}x) = \frac{n_0}{1-  \gamma } [(x_0 + x_2)^{1- \gamma } - (x_0 + x)^{1- \gamma } ], ~~ x_1\leq x\leq x_2.
\end{equation}

A broken power law offers another flexible alternative, characterized by two distinct power-law indices below and above a turnover value $x_0$,
\begin{equation}
N({>}x)\propto \left\{\begin{array}{c}
    x^{-\gamma_1}\quad {\rm for}\,x< x_0\,, \\
    x^{-\gamma_2}\quad {\rm for}\, x\ge x_0\,,
\end{array}    \right.
\end{equation}
where $\gamma_1$ and $\gamma_2$ are the power-law indices, and~$x_0$ is the turning-over value. This model was found to provide a good fit to the cumulative energy distribution of FRB 20201124A~\cite{Xu:2021qdn}.

The Band function, originally proposed for gamma-ray burst spectra \citep{Band:1993eg}, features an exponentially connected broken power law and reads
\begin{equation}
    N({>}x)=\left\{\begin{aligned}
    & Ax^{\hat\alpha}\, {\rm e}^{\left(-x/x_c\right)} \quad &x\le(\hat\alpha-\hat\beta)x_c,\\
       & Ax^{\hat\beta} \, \left[\frac{(\hat\alpha-\hat\beta)x_c}{\rm e}\right]^{\hat\alpha-\hat\beta} \quad &x\ge(\hat\alpha-\hat\beta)x_c.\\
    \end{aligned}\right.
\end{equation}
Here, $\hat\alpha$ and $\hat\beta$ are the power-law indices of the lower and higher energy parts of the distribution, respectively, $x_c$ is a characteristic value for $x$, and~$A$ is the normalization factor. The~Band function has been shown to fit the cumulative energy distribution of FRB 20201124A reasonably~\cite{Zhang:2022rib}.

For sources exhibiting bimodal energy distributions, a~combined log-normal and generalized Cauchy function has been proposed~\cite{Li:2021hpl}
\begin{equation}
N(E) = \frac{N_0}{\sqrt{2\pi}\sigma_E E} \exp\left[\frac{-(\log{E}-\log{E_0})^2}{2\sigma_E^2}\right] +
\frac{\epsilon_{E}}{1+(E / E_0)^{\alpha_E}},
\end{equation}
where $\epsilon_{E}$ is 0 for low-energy part and 1 for high-energy part. Consequently, the~distribution is a pure log-normal function at low energies and a log-normal tail plus a Cauchy term at high energies. Note that here, $N(E)$ denotes the differential distribution, not the cumulative distribution. This composite model has been successfully applied to the burst energy distribution of FRB 20121102A~\cite{Li:2021hpl}.

Unlike non-repeating FRBs, which are observed only once, repeating sources allow the study of the waiting time between successive bursts, a~quantity that is inherently absent for non-repeating sources. The~exponential distribution is the simplest and most fundamental model for the waiting time of repeating FRBs. It arises directly from a Poisson process, where bursts occur randomly and independently at a constant mean rate $r$. The~probability density function is given by
\begin{equation}
    \mathcal{P}(\delta_t \mid r)= r \mathrm{e}^{-r\delta_t}.
\end{equation}
This distribution is memoryless, implying that the probability of a burst occurring in the next instant does not depend on how long one has already waited. Because~of its minimal parameter set, the~exponential model serves as the natural null hypothesis for burst arrival times. In~practice, it has been widely applied to the waiting‑time distributions of many repeating FRBs. For~instance, the~bursts of FRB~20121102A have been shown to be consistent with an exponential distribution after removing short clustering events~\cite{Cruces:2020gmn,Zhang:2021ztz}.

The exponential distribution is insufficient for the full waiting‑time distribution of highly active repeaters, which often show deviations such as excess of short wait times (clustering) or long tails. This limitation naturally motivates the introduction of the Weibull distribution, which extends the exponential distribution by incorporating a shape parameter $k$ that can model temporal clustering ($k<1$) or more regular spacing ($k>1$) relative to the Poisson case. The~Weibull distribution reads~\cite{Opperman:2017kql}
\begin{equation}
    \label{eq:weibull}
    \mathcal{W}(\delta_t \mid k, r)=k \delta_t^{-1}[\delta_t r \Gamma(1+1 / k)]^{k} \mathrm{e}^{-[\delta_t r \Gamma(1+1 / k)]^{k}},
\end{equation}
where $\delta_t$ is the waiting time, $k$ is the shape parameter, $r$ is the mean burst rate, and~$\Gamma$ denotes the gamma function. 
The Weibull distribution generalizes the Poisson process which has a constant rate of bursts. For~the case $k = 1$, the~Weibull distribution reduces to the standard exponential distribution, i.e.,~the Poisson process. 
For values of $k$ different from 1, the~Weibull distribution describes a certain degree of clustering. Specifically, for~$k < 1$, small waiting times are favored relative to the Poisson case, indicating temporal clustering, i.e.,~the occurrence of a burst enhances the probability of another burst in the immediate future. Oppermann~et~al.~\cite{Opperman:2017kql} first proposed to use a Weibull function to describe the distribution of the waiting time and they found the shape parameter $k < 1$ for the waiting time of FRB 20121102A, indicating non-Poissonian~clustering.

Beyond the characterization of energy and waiting-time distributions, the~temporal dynamics of repeating FRBs have been further probed through the analysis of burst sequence fluctuations. For~a given physical quantity $Q$ (e.g., fluence or energy), the~fluctuation over a temporal interval $n$ is defined as $Z_n=Q_{i+n}-Q_i$, where $Q_i$ denotes the value of the $i$-th burst in chronological order. The~normalized fluctuation $z_n = Z_n/\sigma$, with~$\sigma$ being the standard deviation of $Z_n$, has been shown to follow the Tsallis $q$-Gaussian distribution~\cite{Tsallis:1987eu,Tsallis:1998ws}
\begin{equation}
  f(x)=\alpha[1-\beta(1-q)x^2]^{\frac{1}{1-q}},
\end{equation}
where $\alpha$ is the normalization factor, the~parameters $\beta$ and $q$ control the width and sharpness of the peak, respectively. The~$q$-Gaussian
function is a generalization of the Gaussian distribution. It has a much
sharper peak at $x = 0$ and fatter tails at left and right ends than the
Gaussian function. The~deviation from the Gaussian distribution is described by the parameter $q$. In~the limit of  $q \rightarrow 1$, the~$q$-Gaussian function reduces to the Gaussian function with $\mu = 0$ and $\sigma=1/\sqrt{2\beta}$. For~$q>1$, the~distribution develops heavier tails, indicating an enhanced probability of large fluctuations, which is a hallmark of complex, nonlinear systems near a critical state. The~Tsallis $q$-Gaussian distribution has been shown to provide a good fit for the burst fluctuations of various astrophysical and terrestrial phenomena, including earthquakes~\cite{Wang:2015nsl}, soft gamma repeaters (SGRs) \cite{Chang:2017bnb,Sang:2021cjq}, solar gamma-ray flares~\cite{Peng:2023kll}, X-ray flares of gamma-ray bursts~\cite{Wei:2022dbd}, and~repeating FRBs~\cite{Lin:2019ldn,Wei:2021kdw,Sang:2024swg,Gao:2024ekm}.

In summary, the~statistical characteristics of repeating FRBs has evolved considerably beyond simple power-law and Poisson models. The~various empirical functions reviewed above, from~the BPL and Weibull distributions to the composite log-normal/Cauchy form and the $q$-Gaussian fluctuation statistics, each arise from different physical considerations or phenomenological requirements, collectively highlighting the complexity of the underlying emission processes. As~we will see in the following subsections, these models provide essential tools for quantifying the energy, flux, and~temporal properties of individual repeating sources, and~help reveal the dynamical features that distinguish them from other classes of transient astrophysical~phenomena.

\subsection{Energy~Distribution}

Arguably, the most fundamental statistical property of repeating FRBs is their energy distribution, as~it directly reflects the underlying burst generation mechanism. Early studies of repeating sources consistently found that the cumulative energy distribution follows a power-law form over a substantial dynamic range~\cite{Lu:2016fgg,Wang:2019sio,Gourdji:2019lht}. For~example, Lu \& Kumar~\cite{Lu:2016fgg} proposed a universal energy distribution function for repeaters, featuring a power-law index in the range $1.5 < \alpha < 2.2$ and a high-end cutoff $E_{\text{max}}$. Wang \& Zhang~\cite{Wang:2019sio} analyzed bursts from FRB 20121102A observed at multiple frequencies with different telescopes, and~found a power-law index $\alpha$ between 1.6 and 1.8, with~remarkable consistency across epochs and frequency bands, suggesting a universal underlying mechanism for this source. Similarly, based on 41 bursts of FRB 20121102A detected by the Arecibo Telescope, Gourdji~et~al.~\cite{Gourdji:2019lht} reported {a power-law index $1.8 \pm 0.3$ for the cumulative energy distribution, corresponds to a power-law  index  $\alpha = 2.8 \pm 0.3$ for differential distribution, which is much steeper than the previously reported.} However, as~larger and more complete samples have accumulated, it has become clear that a single power law is often inadequate to describe the full dynamic range, and~more complex functional forms are necessary~\cite{Lin:2019ldn,Cheng:2019ykn}.

\textbf{FRB 20121102A}
: The most intensively studied repeating source, FRB 20121102A, provides a prime example. Based on 93 bursts detected with the Green Bank Telescope and 41 bursts detected with the Arecibo Observatory from FRB 20121102A, Lin \& Sang~\cite{Lin:2019ldn} found that the cumulative energy distribution cannot be adequately described by SPL model; instead, it is well fitted by BPL model over the entire energy range. Using 57 bursts from Effelsberg, Cruces~et~al.~\cite{Cruces:2020gmn} found that the cumulative energy distribution is well approximated by a simple power law ($\alpha\sim 1.1$) over the limited range of $\sim$$10^{38}$--$10^{39}$ erg. However, this simple power law fails to provide a global fit to the data over the entire range. This was further confirmed by the much larger samples from FAST. Zhang~et~al.~\cite{Zhang:2021ztz} analyzed 1652 bursts from FRB 20121102A and derived a differential energy index of $\alpha=1.86^{+0.02}_{-0.02}$ for the high-energy regime ($E > 10^{38}$ erg), while lower energies deviated from the power law. Interestingly, they also detected temporal evolution, with~the index changing from $\alpha=1.70^{+0.03}_{-0.03}$ for early bursts to $\alpha= 2.60^{+0.15}_{-0.14}$ for later ones, hinting at possible changes in the emission physics over time. Subsequently, Li~et~al.~\cite{Li:2021hpl} reported a clear bimodal feature in the isotropic-equivalent energy distribution, with~a peak at $4.8 \times 10^{37}$ erg, below~which burst detection is suppressed. They successfully modeled the differential distribution as a combination of a log-normal function at low energies and a generalized Cauchy function at high energies. This bimodality could indicate two distinct burst generation mechanisms, or~alternatively be an observational artifact due to band-limited sensitivity~\cite{Aggarwal:2021qdj}.

\textbf{FRB 20201124A}: Another hyperactive repeater, FRB 20201124A, exhibits similar complexity. Xu~et~al.~\cite{Xu:2021qdn} detected 1863 bursts with FAST and found that a broken power law provides a significantly better fit than a single power law for the cumulative energy distribution. Subsequent work by Zhang~et~al.~\cite{Zhang:2022rib} on 881 bursts from an extremely active episode in September 2021 revealed a clear bimodal differential distribution, well fitted by two log-normal functions, while the cumulative distribution was better modeled by the Band function. Kirsten~et~al.~\cite{Kirsten:2023eqd} further showed that the high-energy tail of FRB 20201124A resembles the distribution of apparently non-repeating FRBs, suggesting that non-repeaters may simply be the rarest events from the same~population.

\textbf{FRB 20220912A}: Zhang~et~al.~\cite{Zhang:2023eui} analyzed 1076 bursts from FRB 20220912A and found that the cumulative energy distribution is well described by a broken power-law function, while the differential energy distribution exhibits a bimodal feature that can be well fitted by the superposition of two lognormal functions. However, Liu~et~al.~\cite{Liu:2025enq} pointed out that this observed bimodality may be partly influenced by band-limited selection effects. They reconstructed the intrinsic energy distribution and found that the low-energy peak disappears, while the high-energy tail steepens. The~intrinsic distribution is better described by a lognormal function with a characteristic energy of $8.13\times10^{37}$ erg for the high-energy component, and~a power-law function with an index of $-1.011\pm0.028$ for the lower-energy part. Their results suggest that the bimodal energy distribution previously reported for FRB 20220912A is not simply an observational artifact, but~may instead originate from the intrinsic radiation mechanism or trigger processes, such as different types of starquakes in magnetars. Pelliciari~et~al.~\cite{Pelliciari:2024uwd} analyzed bursts from FRB 20220912A observed with the Northern Cross radio telescope and found that the cumulative spectral energy distribution (SED) follows a power law with slope $\alpha\sim -1.3$, but~flattens above $E_\nu >10^{31}$ erg Hz$^{-1}$. This flattening, also seen in FRB 20121102A~\cite{Hewitt:2021mex,Jahns:2022evs} and FRB 20201124A~\cite{Kirsten:2023eqd}, indicates that higher-energy bursts occur more frequently than would be expected from a simple power-law extrapolation. In~addition, long-term monitoring with the uGMRT at low radio frequencies shown a break at the lower end of the cumulative energy distribution~\cite{Kumar:2025ypp}.

\textbf{FRB 20240114A}: The recently discovered hyperactive repeater FRB 20240114A has produced the largest burst sample from any FRB source to date~\cite{Shin:2025ybs}. Follow-up long-term monitoring of this source with FAST detected a total of 11,553 bursts between January and August 2024~\cite{Zhang:2025qzn}. Zhang~et~al.~\cite{Zhang:2025qzn} found that its differential energy distribution is better described by bimodal log-normal functions than by a single power law or a single log-normal, indicating the presence of multiple burst populations or distinct emission modes even within a single source. Li~et~al.~\cite{Li:2026wwn} analyzed the same sample and found that none of the single models (SPL, BPL, TPL, or~a Band function) can adequately fit the entire energy range. However, for~subsamples containing more than 50 bursts from single-day observations, the~energy distributions are well described by BPL or TPL models, while their waiting-time distributions are better fitted by a Weibull model. Notably, the~best-fitting BPL parameter $\beta$ remains approximately invariant across the epochs before and after 21 March 2024, with~average values of $1.006\pm0.074$ and $1.236\pm0.183$, respectively. Furthermore, most subsamples from the later epoch exhibit smaller burst rate parameters $r$ compared with those from the earlier epoch, which suggests the presence of two distinct emission~phases.

A systematic comparison of different models applied to the large FAST samples of FRB 20121102A and FRB 20201124A was performed by Sang \& Lin~\cite{Sang:2023zho}. They found that both the bent power-law (BPL) and thresholded power-law (TPL) models can provide good fits to the cumulative energy distributions. Notably, the~BPL index $\beta$ remained nearly invariant across different observing sessions, whereas the median energy parameter $x_b$ varied significantly, suggesting a common emission mechanism whose characteristic energy scale evolves temporally. Although~the TPL model fits the data equally well, its best‑fitting power‑law index $\gamma$ varies significantly with time. In~contrast, the~Band function and SPL models provide much poorer fits, or~even fail, compared to the BPL and TPL~models.

In summary, the~energy distributions of repeating FRBs are far more complex than a simple power law. They routinely exhibit breaks, bimodalities, and~temporal variations that demand sophisticated empirical models. The~growing consensus is that multiple emission modes or distinct burst populations coexist within individual sources, and~that the observed energy distribution is shaped by both intrinsic physical processes and observational selection effects. These empirical constraints provide critical benchmarks for theoretical models of the burst generation and triggering mechanisms in magnetar-like~systems.

\subsection{Flux Density and Fluence~Distribution}

Flux density (or peak flux) measures the instantaneous received power per unit area per unit frequency, while fluence is the time integral of the flux density over the burst duration, representing the total energy per unit area collected during the event. Although~these two quantities emphasize different aspects of the burst emission, with~one capturing peak brightness and the other integrated energy, their statistical distributions have been found to exhibit similar complexity. In~particular, as~with the energy distribution, a~simple power-law function is generally inadequate to describe the full range of fluence or flux density in repeating~FRBs.

For the prototypical repeater FRB 20121102A, early work by Wang \& Yu~\cite{Wang:2016lhy} found that the cumulative distributions of peak flux and fluence from 17 bursts exhibit power-law forms, with~the power-law index $\alpha\sim 1.8$ and $\sim$1.1, respectively. Subsequently, using larger samples of 93 and 41 bursts observed with the Green Bank Telescope and Arecibo Observatory, Lin \& Sang~\cite{Lin:2019ldn} showed that the cumulative distributions of both fluence and flux density are better described by the bent power-law model, consistent with the behavior seen in the energy distribution of the same~source.

For the hyperactive repeater FRB 20220912A, Konijn~et~al.~\cite{Konijn:2024hke} analyzed 696 bursts detected with the Nan{\c{c}}ay Radio Telescope and found that the differential fluence distribution is well fitted by a lognormal function. In~contrast, for~the newly discovered repeater FRB 20240619D, the~cumulative fluence distribution of 249 bursts detected by MeerKAT follows a simple power law above the completeness limit~\cite{Tian:2025zvb}, while a separate analysis of 217 bursts from the same source using the HyperFlash and {\'E}CLAT monitoring programs revealed a power-law break with a flat tail at high energies \citep{Ould-Boukattine:2025zuu}. Such breaks in the fluence/energy distribution are not unique to FRB 20240619D, and~they have also been reported for FRB 20201124A~\cite{Kirsten:2023eqd}, 
FRB 20220912A~\cite{Ould-Boukattine:2024hcc} and FRB 20240114A~\cite{Huang_2025}, strengthening the view that a broken or bent distribution is a common feature of hyperactive repeating~sources.

In summary, the~fluence and flux density distributions of repeating FRBs, like their energy distributions, generally deviate from simple power-law behavior. Lognormal, broken power-law, and~bent power-law models have all been invoked to describe different sources or different energy ranges, suggesting that the observed diversity may reflect a combination of intrinsic emission physics, source-specific activity states, and~observational selection effects. These empirical patterns provide complementary constraints to the energy distribution for understanding the underlying burst generation~mechanism.

\subsection{Waiting-Time~Distribution}

Waiting time characterizes the temporal clustering behavior of repeating FRBs and provides crucial constraints on the burst trigger mechanism. If~bursts were emitted as a simple Poisson process, i.e.,~randomly and independently in time, the~waiting times would follow an exponential distribution with a constant mean occurrence rate~\cite{Wheatland_1998,Wang:2016lhy}. However, observations of ultra-active repeating FRBs have revealed significant departures from this simple picture, pointing to complex, non-Poissonian dynamics. Table~\ref{tab:waitingtime} summarizes FRBs that exhibit multi-peak waiting-time~distributions.

\begin{table}[H]
\caption{FRBs that exhibit multi-peak waiting-time~distributions.}
\label{tab:waitingtime}
\begin{tabularx}{\linewidth}{L L L L L L}
\toprule
\textbf{FRB} & \textbf{Peak 1} & \textbf{Peak 2} & \textbf{Peak 3} & \textbf{Telescope} & \textbf{References} \\
\midrule
20121102A & 3.4 ms & 70 s & -- & FAST & \cite{Li:2021hpl} \\
20121102A & 22 ms & 17.5 s & -- & Arecibo & \cite{Jahns:2022evs}  \\
20201124A & 39 ms & 45.1 s & 162.3 s & FAST & \cite{Xu:2021qdn}\\
20201124A & 51.22 ms & 10.05 s & -- & FAST & \cite{Zhang:2022rib} \\
20201124A & 48 ms & 122 s & -- & uGMRT & \cite{Dudeja:2025tdf} \\
20220912A & 33.4 ms & 67.0 s & -- & Nan\c{c}ay & \citep{Konijn:2024hke} \\
20220912A & 51 ms & 18 s & -- & FAST & \cite{Zhang:2023eui} \\
20240114A & 34 ms & 7.11 s & -- & FAST & \cite{Zhang:2025qzn}\\
20190520B & 32.4 ms & 308.73 s & -- & FAST & \cite{Zhang:2025chw}\\
20200120E & 0.94 s & 23.61 s &  -- & Effelsberg & \cite{Nimmo:2022dra} \\
\bottomrule
\end{tabularx}
\end{table}

\textbf{FRB 20121102A}: The prototypical repeater FRB 20121102A has been the subject of extensive waiting-time analyses across multiple samples. Early work by Wang \& Yu~\cite{Wang:2016lhy}, based on 17 bursts from the Arecibo Observatory and Green Bank Telescope, first suggested that the waiting-time distribution follows a simple power law and is consistent with a non-stationary Poisson process rather than a stationary one. Oppermann~et~al.~\cite{Opperman:2017kql} analyzed the same limited sample and found that a Weibull distribution with shape parameter $k < 1$ provides a better description, indicating temporal clustering, i.e.,~the occurrence of a burst enhances the probability of another burst in the near future. However, the~small sample size ($N = 17$) precluded a definitive conclusion. Based on 41 bursts detected with the Arecibo telescope, Gourdji~et~al.~\cite{Gourdji:2019lht} found that the waiting-time distribution is well described by a log-normal function centered at $\sim$200 s. However, a~subset of bursts shows wait times below 1 s, with~the shortest being 26 ms, in~agreement with earlier evidence for a bimodal distribution~\cite{Li:2019bzv}. Cruces~et~al.~\cite{Cruces:2020gmn} analyzed 57 bursts from Effelsberg telescope and showed that, after~excluding a few closely spaced bursts, the~waiting-time distribution of the remaining events is consistent with a Poisson process, namely, with~a Weibull shape parameter $k$ consistent with unity. Furthermore, using larger samples of 93 and 41 bursts obtained with the Green Bank Telescope and the Arecibo Observatory, respectively, Lin \& Sang~\cite{Lin:2019ldn} found that the cumulative waiting-time distribution is well fitted by a bent power-law model, whereas a simple power law fails to provide an adequate fit to the~data.

The picture changed significantly with the advent of much larger samples from FAST telescope. Li~et~al.~\cite{Li:2021hpl} analyzed 1652 bursts from FRB 20121102A and found that the differential waiting-time distribution is dominated by a log-normal component peaking at 70 s (or 220 s for the high-energy subsample), consistent with a stochastic process. They also identified a secondary peak at $\sim$3.4 ms, most likely arising from the substructure of individual bursts rather than true inter-burst waiting times. Using the same dataset, Zhang~et~al.~\cite{Zhang:2021ztz}  further demonstrated that the waiting-time distribution is well fitted by a Weibull function with $k=0.72$, confirming the presence of temporal clustering. However, when waiting times shorter than 28 s are excluded, the~remaining burst sequence becomes roughly consistent with a Poisson process, suggesting that clustering is primarily driven by short inter-burst intervals. Based on 133 bursts observed by Arecibo, Aggarwal~\cite{Aggarwal:2021quq} also found a bimodal feature in the waiting-time distribution, which was later further confirmed by a much larger sample~\cite{Jahns:2022evs}. Jahns~et~al.~\cite{Jahns:2022evs} analyzed 849 bursts from FRB 20121102A observed with Arecibo and found a bimodal waiting-time distribution, with~clusters on a short timescale of 22 ms and a long timescale of 17.5 s. The~latter is consistent with a non-stationary Poisson process, in~which the burst rate varies between observing sessions but follows Poisson statistics within each session. They confirmed that two-parameter Weibull model is not needed to explain the wait-time distribution within a single~observation. 

\textbf{FRB 20201124A}: The hyperactive repeater FRB 20201124A exhibits even richer multimodal structure. Xu~et~al.~\cite{Xu:2021qdn} detected 1863 bursts with FAST and found that the differential waiting-time distribution can be well described by the superposition of three log-normal components, with~peaks at 39 ms, 45.1 s, and~162.3 s. In~a separate study of 881 bursts from an extremely active episode from this source, Zhang~et~al.~\cite{Zhang:2022rib} reported a double-peak distribution modeled by two log-normal functions peaking at 51.22 ms and 10.05 s, respectively. Notably, the~second peak time was shorter than that observed in a previous active episode~\cite{Xu:2021qdn}, suggesting that the location of this peak is tied to the source's global activity level. Dudeja~et~al.~\cite{Dudeja:2025tdf} studied FRB 20201124A with uGMRT and also reported a bimodal waiting‑time distribution, with~log‑normal peaks at $\sim$48 ms and $\sim$122 s, and~detected independent burst pairs separated by as little as $\sim$17 ms. Their multi‑frequency observations revealed that higher‑frequency emission ceased earlier than lower‑frequency activity, and~they found sub‑second temporal offsets ($\sim$1.1 s) between bursts in different bands, suggesting intrinsic or propagation‑related effects. These results highlight the complex, frequency‑dependent, and~multi‑timescale nature of repeating~FRBs.

A systematic comparison of FRB 20121102A and FRB 20201124A by Sang \& Lin~\cite{Sang:2023zho} revealed that the waiting-time distribution parameters evolve with time. For~both sources, the~waiting time in each observing session can be well fitted by the exponential distribution, with~the mean occurrence rate $r$ varying significantly across different observing sessions, underscoring the non-stationary nature of the burst emission process and the need for time-dependent modeling~approaches.

\textbf{FRB 20220912A}: Similar bimodality is also observed in FRB 20220912A. Konijn~et~al.~\citep{Konijn:2024hke} analyzed 696 bursts from FRB 20220912A detected with the Nan\c{c}ay radio telescope, and~found a bimodal waiting-time distribution with peaks at 33.4 ms and 67.0 s. While the burst rate within individual observations appears Poissonian, the~full dataset is better modeled by a Weibull distribution, implying that the emission process operates on multiple timescales, with~very short inter-burst intervals possibly arising from within a single active episode, and~longer intervals separating distinct activity windows. Zhang~et~al.~\cite{Zhang:2023eui} detected 1076 bursts from FRB~20220912A with FAST and found a bimodal waiting-time distribution with peaks at $\sim$51~ms and $\sim$18~s. The~distribution of the main peak ($\sim$18~s) is well fitted by either a log-normal or an exponential model. The~secondary peak ($\sim$51\,ms) resembles that seen in other repeaters such as FRB 20121102A~\cite{Li:2021hpl} and FRB 20201124A~\cite{Xu:2021qdn,Zhang:2022rib,Dudeja:2025tdf}, suggesting a common intrinsic~timescale.

\textbf{FRB 20240114A}: The most recently discovered hyperactive repeater, FRB 20240114A, also exhibits a bimodal waiting-time distribution. Using the uGMRT, Panda~et~al.~\cite{Panda:2024ivc} detected 167 bursts from FRB~20240114A and identified a bimodal waiting-time distribution. The~short-waiting component ($<$1\,s) follows a Weibull distribution with $k=0.63$, whereas the long-waiting component ($>$1\,s) is described by a log-normal distribution with $\sigma=1.28$. Notably, the~log-normal width for long waits is comparable to that seen in other hyperactive repeaters. In~addition, based on 11,553 bursts from the same source observed with FAST, Zhang~et~al.~\cite{Zhang:2025qzn} also report a clearly bimodal waiting-time distribution, with~two log-normal peaks at 34~ms and 7.11~s.

Similar bimodal behavior has also been reported in several other repeaters, indicating that it may be a common property of this population. Notable examples include FRB 20190520B, with~peaks at 32.4 ms and 308.73 s~\cite{Zhang:2025chw}, and~FRB 20200120E, with~peaks at 0.94~s and 23.61 s~\cite{Nimmo:2022dra}. More recently, FRB 20240619D has been shown to display temporal clustering, with~its waiting-time distribution well described by a Weibull function with shape parameter \(k<1\) across a wide frequency range~\cite{Shaji:2026mdn}, consistent with that found in FRB 20121102A~\cite{Opperman:2017kql,Zhang:2021ztz}.

In summary, the~waiting-time distributions of repeating FRBs robustly deviate from a simple Poisson process. They are characterized by multimodal log-normal or Weibull forms, reflecting the coexistence of multiple timescales in the burst generation process, from~millisecond-scale substructures to minute- to hour-scale activity cycles. The~observed non-stationarity and temporal clustering strongly point toward complex triggering mechanisms, such as self-organized criticality in magnetar crusts, rather than a simple independent-event process. These temporal signatures provide complementary constraints to the energy distributions reviewed above, and~together they form the basis for the nonlinear time-series analyses that we turn to~next.

\subsection{Time Series~Analysis}

Beyond the characterization of energy and waiting-time distributions, the~temporal dynamics of repeating FRBs have been extensively studied through nonlinear time-series analysis. These approaches aim to uncover the underlying physical processes by quantifying the randomness, memory, and~scale-invariant properties of burst sequences. Several nonlinear dynamical statistics have been employed in this context, including the Pincus index (PI), the~largest Lyapunov exponent, the~Hurst exponent, and~the Tsallis $q$-Gaussian distribution. Together, these tools provide complementary insights into whether the burst emission is driven by stochastic processes, deterministic chaos, or~self-organized~criticality.

Zhang~et~al.~\cite{Zhang:2023fmn} pioneered the application of nonlinear dynamics to FRBs by introducing the PI and the largest Lyapunov exponent to quantify the randomness and chaoticity of burst sequences. Through a systematic investigation of two active repeaters, FRB 20121102A and FRB 20190520B, in~the time–energy bivariate space, they found that the burst behaviors exhibit high randomness and low chaos, closely resembling Brownian motion. This stands in sharp contrast to other transient phenomena such as pulsars, earthquakes, and~solar flares. Moreover, they found no correlation between waiting times and the corresponding energy changes, suggesting that FRB emission does not exhibit the time–energy clustering observed in seismic~events.

Subsequent studies have revealed the presence of long-range memory effects in repeating FRBs. Using the two largest FAST samples from FRB 20121102A and FRB 20201124A, Wang~et~al.~\cite{Wang:2023wcb} found that the burst-rate structures exhibit coherent growths over timescales ranging from minutes to about an hour, as~quantified by the Hurst exponent. They further conducted cellular automaton simulations and interpreted the observed memory as evidence for a self-organized criticality system, pointing toward a neutron star crustal failure model as a plausible triggering mechanism. Wang~et~al.~\cite{Wang:2023sjs} extended this analysis through a systematic conditional waiting-time study based on four FAST datasets, FRB 20121102A, two active periods of FRB 20201124A, and~FRB 20220912A, and~found a universally present temporal clustering memory effect where short waiting times tend to follow short ones, and~long waiting times tend to follow long~ones.

A more comprehensive analysis of memory effects was performed by Sang \& Lin~\cite{Sang:2024swg} using four FAST samples from three extremely active repeaters (FRB 20121102A, FRB 20201124A, and~FRB 20220912A). They employed the Hurst exponent, the~Pincus index, and~the Tsallis $q$-Gaussian distribution to study the statistical properties of energy and waiting-time time series. The~Hurst exponents for both quantities across all samples were significantly greater than 0.5, confirming the presence of persistent long-range correlations. While the PI for waiting times was close to 1.0, it still deviated from complete randomness at a confidence level exceeding 3$\sigma$, indicating subtle yet significant structure. In~contrast, the~randomness of energy time series showed notable source-to-source variations, with~FRB 20121102A exhibiting a significant deviation from random organization, while the other three samples were consistent with a fully random energy distribution. These analytical approaches have also been extended to magnetar bursts, where similar memory effects and dynamical stability have been quantified~\cite{Sang:2024upe}.

Zhang~et~al.~\cite{Zhang:2026jwa} further examined the hyperactive repeater FRB 20240114A with 11,553 bursts from FAST and uncovered a more nuanced, scale‑dependent memory. On~short timescales (seconds to hours), the~waiting‑time distribution is a superposition of three exponentials (3Exp), and~the Hurst exponent is near 0.5, indicating random behavior. On~longer timescales (weeks to months), a~power‑law tail emerges, and~the Hurst exponent exceeds 1, {revealing the existence of long-term memory}. The~transition occurs at about one hour, consistently seen in both coherence analysis and R/S analysis. The~energy distribution also shows a waiting‑time‑dependent Gutenberg‑Richter index, with~shallower slopes for short‑wait bursts. This work demonstrates that memory in repeating FRBs is not monolithic but depends on the observational timescale, providing a new benchmark for burst~models.

Xu~et~al.~\cite{Xu:2026ebe} carried out a systematic comparison of repeating FRBs with magnetar flares, pulsar glitches, earthquakes, and~solar flares using the Pincus index (PI) and the largest Lyapunov exponent (LE) as two complementary diagnostics of stochasticity and chaos. Mapping all sources onto the Pincus--Lyapunov diagram (PLD), they found that repeating FRBs occupy a distinct region characterized by high PI ($\sim$0.90) and low LE ($\sim$0.09), clearly separated from magnetar flares and pulsar glitches. Statistical tests based on energy distance and permutation confirm that the FRB population is intrinsically distinct from both magnetar flares and glitches. Notably, for~the hyperactive repeater FRB 20240114A, the~(PI, LE) values remain stable across different observing sessions despite burst‑rate variations of orders of magnitude, indicating that the statistical signature is intrinsic rather than driven by short‑term activity changes. This distinct localization suggests that repeating FRBs are not simple radio analogs of magnetar X‑ray flares or starquake‑driven glitches, but~rather are governed by a highly stochastic, weakly chaotic~process.

The correlation between burst occurrence times and energies provides a powerful diagnostic for the physical mechanism of repeating FRBs. Totani \& Tsuzuki~\cite{Totani:2023dxo} performed a comprehensive correlation-function analysis on nearly 7000 bursts from three active repeaters (FRB~20121102A, FRB~20201124A, and~FRB~20220912A) in the two-dimensional time–energy space. They discovered a clear power-law signal in the correlation function at time separations below $\sim$1~s, with~an aftershock rate decaying as $\propto (\Delta t + \tau)^{-p}$ where $p \sim 1.5\text{--}2.5$, matching the Omori–Utsu law of earthquakes. Moreover, they found that the branching ratio (the expected number of aftershocks per event) is about $0.1\text{--}0.6$, similar to that of seismic events, and~that there is little correlation between energy release and time delay. In~sharp contrast, solar flares showed significantly different correlation functions, indicating that repeating FRBs are fundamentally earthquake-like phenomena driven by starquakes in neutron-star crusts rather than by magnetic reconnection analogous to solar~flares.

Complementary to the subsecond-scale earthquake analogy, the~longer temporal behavior of repeating FRBs reveals a different scaling picture. Du~et~al.~\cite{Du:2023gkm} analyzed the waiting-time distributions of the FAST samples (FRB 20121102A, FRB 20201124A, and~FRB 20220912A) for intervals longer than $\sim$1 s and uncovered a unified scaling law. This behavior is nearly identical to that observed in solar flares, rather than earthquakes. Based on this dichotomy, they proposed a dual analogy: subsecond bursts resemble earthquake aftershocks (consistent with crustal fracture), while longer intervals mimic solar-flare dynamics, possibly reflecting convective magnetic energy transport within the neutron star. This suggests that the $\sim$1~s threshold marks the dynamical timescale of the neutron-star crust, offering a unified view of FRB temporal~complexity.

Kimpson \& O'Leary~\cite{Kimpson:2026xdu} applied $\varepsilon$‑machine reconstruction, a~method that extracts the minimal predictive model from a time series, to~the waiting‑time sequences of three repeating FRBs. They found that the two FRBs observed by FAST (FRB 20121102A and FRB 20201124A) each carry about one bit of temporal memory, meaning their burst sequences are not random but require one bit for optimal prediction. This memory operates between observing sessions, not within them: for FRB 20121102A the order of sessions matters, while for FRB 20201124A the signal comes from differences in activity rates across sessions. The~FRB observed by CHIME (FRB 20220912A) showed no detectable memory, but~the authors caution that CHIME's short transit windows could suppress such structure. This work complements earlier complexity measures (e.g., Hurst exponent, Pincus index) by providing a model‑free, minimal generative description of burst~timing.

In addition to analyzing burst sequences directly, the~temporal fluctuations of observed physical quantities have also been investigated. Using 93 and 41 bursts from FRB 20121102A observed by the Green Bank Telescope and Arecibo Observatory, Lin \& Sang~\cite{Lin:2019ldn} found that the probability density functions of fluctuations of fluence, flux density, and~total energy are well described by the Tsallis $q$-Gaussian distribution, with~best-fitting $q \sim2$ that remains constant across different temporal scales, which is a hallmark of scale invariance. This behavior is nearly identical to that observed in earthquakes~\cite{Wang:2015nsl} and magnetar bursts~\mbox{\cite{Chang:2017bnb,Sang:2021cjq}}. Based on these statistical similarities, they explicitly proposed that the origin of repeating FRBs may involve starquakes on compact stars, analogous to earthquakes on Earth, providing an early statistical-physics clue to the triggering mechanism. The~$q$-Gaussian nature and scale invariance of temporal fluctuations were later confirmed using much larger samples. Sang \& Lin~\cite{Sang:2024swg} analyzed four datasets with 1652, 1863, 881, and~1076 bursts, respectively, from~three extremely active repeaters and found that the fluctuations of both energy and waiting time are universally well fitted by the Tsallis $q$-Gaussian distribution, with~$q$ values independent of temporal interval, consistent with the scale-invariance property of SOC~systems.

Gao \& Wei~\cite{Gao:2024ekm} extended the scale-invariance analysis to FRB~20201124A and compared its properties with those of the glitching pulsar PSR~B1737-30. They confirmed that for this repeating FRB, the~fluctuations of energy, peak flux, duration, and~waiting time all follow the Tsallis $q$-Gaussian distribution, with~$q$ values remaining nearly constant across different temporal interval scales, a~hallmark of scale invariance previously established for FRB~20121102A. Remarkably, the~same $q$-Gaussian behavior was also found for the fluctuations of glitch size and waiting time in PSR~B1737--30, implying a physical connection between repeating FRBs and pulsar glitches, both naturally explained by self-organized criticality. Furthermore, a~comparison between low-energy (from FAST) and high-energy bursts revealed that the $q$ values for duration and waiting time differ significantly between the two samples, suggesting that high-energy bursts may originate from a distinct emission mechanism or region within the same~progenitor.

In summary, nonlinear time-series analyses have firmly established that repeating FRBs exhibit high randomness, persistent memory effects, and~scale-invariant temporal fluctuations. These dynamical signatures, particularly their resemblance to earthquakes and magnetar bursts, provide compelling evidence that repeating FRBs are driven by complex, correlated processes in highly magnetized neutron star crusts, and~they offer powerful constraints for theoretical models of burst triggering and emission~mechanisms.

\subsection{Summary for Repeating FRBs}

The statistical properties of repeating FRBs, as~synthesized from extensive observations of hyperactive sources, reveal a picture of striking complexity that defies simple phenomenological descriptions. Across the energy, fluence, and~flux density domains, distributions routinely deviate from single power-law behavior. Instead, they are more accurately described by bent power-law, thresholded power-law, broken power-law, or~bimodal log-normal functions, depending on the source and the energy range considered. This diversity of functional forms points to the existence of characteristic energy scales and, in~several cases, suggests the coexistence of multiple distinct emission modes or burst populations within a single~source.

Temporally, the~emission from repeating FRBs is characterized by pronounced non-Poissonian clustering. Waiting-time distributions are typically multimodal, well described by Weibull or multi-peaked log-normal functions, reflecting the operation of multiple timescales, from~millisecond-scale substructures within individual bursts to minute- to hour-scale activity cycles. Moreover, the~distribution parameters are found to evolve with time, underscoring the intrinsically non-stationary nature of the burst generation~process.

Nonlinear time-series analyses have uncovered deeper dynamical signatures. Repeating FRBs exhibit persistent long-range memory effects, as~quantified by Hurst exponents significantly greater than 0.5, together with scale-invariant temporal fluctuations that follow the Tsallis $q$-Gaussian distribution, which is a hallmark of complex systems near a critical state. These dynamical characteristics---high randomness, low chaos, memory and scale invariance---bear a striking resemblance to those observed in terrestrial earthquakes and magnetar~bursts.

The statistical fingerprints of repeating FRBs strongly point toward a physical origin rooted in self-organized criticality, most plausibly manifested as starquakes in the highly magnetized crusts of neutron stars. While the exact triggering mechanism remains to be pinpointed, the~unified empirical framework established by these statistical analyses provides indispensable constraints for any viable theoretical model aiming to explain the burst generation and emission physics of these enigmatic~sources.

\section{Periodicity}\label{sec:periodicity}

\subsection{Observational~Discoveries}

The study of periodicity in FRBs is crucial, as~it provides direct insight into their underlying physical nature. Periodic activity, whether indicative of a rotating neutron star or orbital motion in a binary system, offers tight constraints on the central engine powering these bursts. Moreover, it enables astronomers to predict active windows, rigorously test emission models, and~utilize these bursts as sensitive probes of extreme magnetic environments. Ultimately, periodicity elevates FRBs from random, transient signals to predictable astrophysical laboratories. Given this profound significance, considerable effort has been devoted to searching for periodicity in FRBs. Table~\ref{tab:sources} summarizes the representative periodicities detected in FRB sources to~date.

\begin{table}[H]
\small
\caption{Periodic FRB sources and their observed~periodicities.}
\label{tab:sources}
\begin{tabularx}{\linewidth}{L l L l L}
\toprule
\textbf{FRB} & \textbf{Period} & \textbf{Type} & \textbf{Method \textsuperscript{1} 
} & \textbf{Reference} \\
\midrule 
20180916B & $16.35$ days ($5\sigma$) & Burst arrivals & PFA, H-test& \cite{CHIMEFRB:2020bcn,Pleunis:2020vug,Gopinath:2023fty,Pastor-Marazuela:2020tii,Bethapudi:2022mfe} \\
20121102A & $159.3$ days ($>$5$\sigma$) & Burst arrivals & LSP & \cite{Braga:2024ntx,Rajwade:2020uat,Cruces:2020gmn} \\
20121102A & 4.605 days ($\sim$9$\sigma$) & Burst arrivals & PBA & \cite{Li:2024ckk} \\
20201124A & $26.24$ days ($5.9\text{--}34\sigma$) & RM variation & LSP & \cite{Xu:2025hal} \\
20201124A $^a$ & $\sim$1.7 s ($7.3\sigma$) & Burst arrivals & PFA & \cite{Du:2025lyg} \\
20220529A & $\sim$200 day ($3.1\text{--}4.1\sigma$) & RM variation & LSP, PFA & \cite{Liang:2025tln} \\
20240209A & $\sim$126 days ($>$3$\sigma$) & Burst arrivals & ACF, LSP & \cite{Pal:2025qxc}\\
20240114A & multi-timescale & Burst arrivals & TDA, LSP, ACF & \cite{Zhou:2025acx}\\
20191221A $^b$ & $216.8$ ms ($6.5\sigma$) & Substructure & FFT, Rayleigh test & \cite{CHIMEFRB:2021fvq}\\
20230708A $^c$ & $7.267$ ms ($1.77\sigma$) & Substructure  & ACF & \cite{Dial:2024ljo}\\
\bottomrule
\end{tabularx}
\noindent\small{Note: $^a$ The 1.7 s signal was questioned by the reanalysis of Gazith \& Zackay~\cite{Gazith:2025gym}. $^b$ This event was recently identified as a series of pulses from the Galactic pulsar PSR J0248+6021, rather than an FRB~\cite{CHIMEFRB:2026mjq}. $^c$ Marginal candidate: the 7.267 ms signal in FRB 20230708A is reported at only $1.77\sigma$ and is not considered a secure periodicity. \textsuperscript{1} Periodicity search methods: Lomb--Scargle periodogram (LSP), Phase-folding algorithm (PFA), Phase-folding probability binomial analysis (PBA), Autocorrelation function (ACF), Time-differencing algorithm (TDA), Fast Fourier transform (FFT), H-test and Rayleigh test.}
\end{table}

\textbf{FRB~20180916B}: This is the first FRB source found to exhibit robust periodic activity. The~CHIME/FRB collaboration reported a periodicity of 16.35 days in its burst activity, with~an active window of $\sim$5 days during each cycle~\cite{CHIMEFRB:2020bcn}. Subsequent analysis from the LOFAR samples further confirmed this periodicity (16.33 days), but~with a relatively narrower active window ($\sim$4.3 days) \cite{Gopinath:2023fty}. A~remarkable feature of FRB~20180916B is its chromatic (frequency-dependent) activity cycle~\cite{Pleunis:2020vug}. Multi-frequency observations revealed that bursts at higher frequencies ($400-800$~MHz, CHIME band) appear earlier in the activity cycle, while lower-frequency bursts ($110-188$~MHz, LOFAR band) appear several days later~\cite{Gopinath:2023fty}. This chromatic behavior was corroborated by simultaneous Apertif and LOFAR observations spanning a broad frequency range (120--1400~MHz)~\cite{Pastor-Marazuela:2020tii}. It was further confirmed by Effelsberg radio telescope measurements at 4--8~GHz, which detected an active window of 1.35 days that precedes the CHIME activity peak by 3.6 days~\cite{Bethapudi:2022mfe}. {In addition, Pastor-Marazuela~et~al.~\cite{Pastor-Marazuela:2020tii} found that the activity windows are narrower at higher frequencies, which is opposite to the predictions of the simple stellar-wind free–free absorption models. Such frequency-dependent behavior strongly disfavors simple free–free absorption in a stellar wind, and~ binary models more generally remain viable only if they can reproduce the observed frequency-dependent window widths.}

Polarization observations of FRB~20180916B have provided additional constraints on its periodic origin. The~rotation measure (RM) of this source exhibits variations with time, but~remarkably, these RM variations do not follow the 16.33-day activity cycle~\cite{Gopinath:2023fty}, suggesting that the magnetoionic environment and the burst emission geometry may be decoupled to some~extent.

\textbf{FRB~20121102A}: As the first detected repeating fast radio burst, FRB~20121102A is one of the most active sources known to date. This source exhibits periodic activity on two distinct~timescales.

\textit{Long-term periodicity}: A cyclic activity pattern has been revealed by long-term monitoring of this source, with~a period of approximately 157 days derived from Lovell telescope data~\cite{Rajwade:2020uat}, and~161 days from Effelsberg telescope data~\cite{Cruces:2020gmn}. Using the Effelsberg radio telescope at L-band, Braga~et~al.~\cite{Braga:2024ntx} performed a detailed periodicity analysis and derived a consistent period of 159.3 days with two different models, along with an activity phase of approximately 53\%. The~activity window exhibits a Gaussian-like profile, with~the burst detection rate peaking at the center of the window and gradually declining toward the edges. Crucially, the~periodicity found through both binary detection models and normalized rates models is consistent across L-band and S-band datasets, indicating that the $\sim$159-day modulation is intrinsic to the source rather than an observational artifact. However, the~C-band sample remains too small to confirm the chromatic behavior of the activity window for this~source.

\textit{Short-term periodicity}: An intriguing additional short-term periodicity in FRB 20121102A was reported from FAST observations by Li~et~al.~\cite{Li:2024ckk}. Alongside the previously known~157-day period, they identified a candidate period of 4.605 days, which becomes more significant when only higher‑fluence bursts (above $10^{38}$ erg) are considered. This finding suggests that the periodic behavior of FRB 20121102A may be multi‑scale and burst‑energy dependent, adding further complexity to its activity~interpretation.

\textbf{FRB 20240114A}: The hyperactive repeater FRB 20240114A also exhibits remarkable multi-timescale periodicity~\cite{Zhou:2025acx}. On~short timescales, FAST detected three candidate periodic signals in the time-of-arrival data at $0.673$, $0.635$ and $0.536\,\mathrm{s}$ ($3.2\sigma\text{--}6\sigma$ significance) across two independent observational epochs. On~longer timescales, a~dominant period of $143.40 \pm 7.19\,\mathrm{d}$ with $5.2\sigma$ significance and a secondary period of $73.60 \pm 2.45\,\mathrm{d}$ with $3.3\sigma$ significance were identified. In~addition, burst time series reveal transient quasi‐periodic oscillations at hundreds of hertz ($3.4\sigma$ and $3.7\sigma$) as well as periodic burst trains with periods of several to tens of milliseconds ($3\sigma\text{--}3.9\sigma$), all of which are short‐lived. Despite this rich set of periodic signatures, no definitive spin period of the source is identified, placing strong constraints on progenitor models and suggesting that the observed modulations are intermittent and likely arise from multiple physical processes operating on distinct timescales. Long-term monitoring of FRB 20240114A with the GMRT shows that this source exhibits chromaticity in its burst activity~\cite{Kumar:2024svu}, similar to that observed in FRB 20180916B~\cite{Pleunis:2020vug,Pastor-Marazuela:2020tii}.

\textbf{FRB~20201124A}: The highly active repeating source FRB 20201124A stands out as an exceptionally fertile ground for periodicity analysis, showing two distinct types of periodic~activity.

\textit{Faraday RM periodicity}: Xu~et~al.~\cite{Xu:2025hal} analyzed a large sample of bursts (more than 3000) from FRB 20201124A observed with the FAST telescope over nearly one year and detected a $26.24 \pm 0.02$ day periodicity in the Faraday RM. The~RM signal remains phase‑connected coherently across approximately 14 cycles, with~detection significances between $5.9\sigma-34\sigma$ depending on the underlying assumptions. This finding strongly supports the binary hypothesis, since the observed RM modulation is consistent with the orbital motion of the burst source through the magnetoionic medium of a binary companion~\cite{Xu:2025hal}.

\textit{Second-scale burst periodicity}: In an independent study, Du~et~al.~\cite{Du:2025lyg} reported the discovery of a $\sim$1.7\,s ($7.3\sigma$) periodicity in the burst arrival times of FRB~20201124A. From~49~days of observations by FAST that yielded over 2800 bursts, the~source exhibited a clear periodic signal on two specific days: $1.706015(2)$~s on MJD~59310 and $1.707972(1)$~s on MJD~59347. Notably, no such periodicity was detected during the remaining 47 days, implying that the 1.7 s signal may be transient or condition-dependent. The~measured period derivative of $6.14 \times 10^{-10}$~s~s$^{-1}$ implies a surface magnetic field strength of $1.04 \times 10^{15}$~G and a spin-down age of only 44 years for the central engine, strongly pointing to a young magnetar origin for FRB~20201124A~\cite{Du:2025lyg}. {However, a~subsequent reanalysis by Gazith \& Zackay~\cite{Gazith:2025gym} of the same dataset found no statistically significant second-scale periodicity. They argue that the claimed detection is sensitive to the treatment of bursts with short waiting times and to the estimation of the null distribution of the periodicity search. The~1.7 s signal claimed by~\cite{Du:2025lyg} should therefore currently be regarded as a contested candidate rather than a confirmed detection.}

\textbf{FRB 20220529A}: This is another source that exhibits possible periodicity in its RM variation. Based on nearly three years of FAST observations of FRB 20220529A, Liang~et~al.~\cite{Liang:2025tln} report a possible~200-day periodicity in the RM evolution, with~significances of $4.1\sigma$ (LSP) and $3.1\sigma$ (PFA). The~RM exhibits a dramatic increase followed by a rapid recovery on a timescale of about one week. The~time-of-arrival also exhibits a similar periodic feature, but~with much less significance. If~confirmed, such RM periodicity would provide a novel probe of binary orbital motion or a precessing magnetized environment, independent of burst arrival-time~periodicity.

\textbf{FRB 20240209A}: This repeating FRB resides in the outskirts of a quiescent elliptical galaxy at $z=0.1384$, marked by a large offset of $40\pm5$~kpc from the host galactic center. As~reported by Pal~\cite{Pal:2025qxc}, a~$\sim$126-day periodicity is detected in this source, with~statistical significance confirmed by bootstrap analysis. The~detected periodicity favors progenitor models invoking binary systems (a compact object orbited by a stellar companion) or precessing/rotating old magnetars, while the quiescent and old stellar environment strongly disfavors young magnetar~interpretations.

\textbf{FRB 20191221A}: {All of the FRBs mentioned above are repeating sources. However, the~apparently non-repeating burst FRB~20191221A was reported by the CHIME/FRB Collaboration~\cite{CHIMEFRB:2021fvq} as a multicomponent event with a total duration of $\sim$3~s and at least nine distinct subcomponents. A~timing analysis revealed a significant periodicity of $216.8(1)$~ms between the components, with~a confidence level of $6.5\sigma$. The~measured scattering timescale ($\tau_s = 340$~ms at 600~MHz) and the narrow intrinsic component widths ($\sim$4~ms) were argued to be consistent with a neutron-star progenitor, favoring emission from the magnetosphere over more distant emission regions. Two other bursts, FRB~20210206A and FRB~20210213A, exhibit possible periodicities of $\sim$2.8~ms and $\sim$10.7~ms, respectively, but~their significances remain inconclusive. However, the~FRB~20191221A detection has since been reidentified as a series of pulses from the Galactic pulsar PSR~J0248+6021, caused by a CHIME calibration and beam-pointing error that produced an incorrect localization and an overestimated dispersion measure~\cite{CHIMEFRB:2026mjq}. This event therefore no longer provides evidence for periodic magnetospheric emission from an extragalactic FRB, and~highlights the need to rule out Galactic pulsar contamination and calibration systematics in periodicity searches.}

\textbf{FRB 20230708A}: Alongside FRB~20191221A, the~apparently non-repeating FRB~20230708A likewise exhibits a complex temporal structure suggestive of potential periodic modulation. Dial~et~al.~\cite{Dial:2024ljo} reported this multicomponent event based on ASKAP observations, with~a duration of $\sim$26.44~ms and comprising at least 11 distinct subcomponents. A~candidate periodicity of $7.267$~ms is identified, albeit with marginal significance ($1.77\sigma$). The~scattering timescale ($\tau_s = 0.17$~ms) is consistent with a neutron-star progenitor. Polarimetric data reveal structured position angle sweeps and significant circular polarization; however, propagation effects are not favored as the primary cause. Overall, the~properties are more consistent with a magnetar origin than with a millisecond~pulsar.
 
\subsection{Period Detection Methods and Null~Results}

Several statistical techniques have been employed to search for periodicities in FRB datasets. Since different methods are sensitive to distinct signal characteristics, it is standard practice to apply multiple algorithms and cross‑check their results. This multi‑method strategy helps mitigate false positives and increases confidence in any detected periodic signal. The~most common methods~include the following:
\begin{itemize}
    \item\textbf{Lomb--Scargle periodogram (LSP):} This is a frequency-domain technique that estimates the power spectrum by fitting sinusoidal models at each trial frequency, making it particularly suited for irregularly sampled time series. It has been extensively applied to both FRB~20121102A~\cite{Cruces:2020gmn,Braga:2024ntx} and FRB~20201124A~\cite{Xu:2025hal}, and~to search for the long-timescale periodicity of FRB 20240114A~\cite{Zhou:2025acx}. The~method accommodates the sparsity of FRB detection datasets by treating detections either as binary flags (1 for detection, 0 for non-detection) or via normalized burst~rates.

    \item\textbf{Phase-folding algorithm (PFA):} This is a time-domain technique that folds burst arrival times modulo a trial period and searches for significant phase clustering. It has been successfully applied to detect the first discovered periodicity in FRB~20180916B~\cite{CHIMEFRB:2020bcn}, and~the 1.7 s periodicity in FRB~20201124A~\cite{Du:2025lyg}. The~method is particularly sensitive to periodic signals with narrow duty~cycles.
    
    \item\textbf{Phase-folding probability binomial analysis (PBA):} This method extends phase-folding algorithm by jointly searching over the period $T$, central phase $P$, and~active window half-width $\Delta P$. It counts bursts within each trial window and evaluates significance via a binomial test with false-alarm probability correction. PBA is particularly sensitive to quasi-periodic signals with narrow duty cycles, which are often missed by traditional frequency-domain techniques. This method was applied to FRB~20121102A and recovered the known $\sim$157-day period and revealed a new $4.605$-day candidate period~\cite{Li:2024ckk}.

    \item\textbf{Time-differencing algorithm (TDA):} This is a time-domain method that evaluates periodicity by analyzing time differences between events across a sliding window, making it particularly efficient for sparse and irregularly distributed time-of-arrival data. It has been successfully applied to search for short-timescale periodicities in FRB 20240114A, identifying several candidate signals with periods of $\sim$0.5--0.7 s~\cite{Zhou:2025acx}.

    \item\textbf{Autocorrelation function (ACF):} This time-domain method measures the correlation between a time series and itself at different time lags, identifying periodicities through significant peaks in the correlogram. Unlike frequency-domain techniques, ACF method does not assume a sinusoidal waveform, making it robust for detecting non-sinusoidal or asymmetric burst activity patterns. The~method has been successfully applied to uncover the $\sim$126-day periodicity in the activity of FRB~20240209A~\cite{Pal:2025qxc}, and~the $7.267$ ms periodicity in the substructure of 20230708A~\cite{Dial:2024ljo}.

    \item\textbf{H-test and Rayleigh test:} These methods are designed to detect periodic modulations in event arrival times and have been employed in short-timescale periodicity searches, particularly for significance assessment~\cite{Zhang:2018jux,CHIMEFRB:2021fvq,Du:2023qpa}. They have been successfully used to uncover the subsecond periodicity in FRB~20191221A~\cite{CHIMEFRB:2021fvq}.
\end{itemize}

Despite the expectation that rotating magnetized neutron stars should exhibit rotational periodicity in their emission, comprehensive searches have failed to detect clear periodic signals in the millisecond-to-second range from most repeating FRBs. Du~et~al.~\cite{Du:2023qpa} performed an exhaustive search for short-timescale periodicity (from 1 ms to 1000 s) in four active repeating FRBs (20121102A, 20200120E, 20201124A, and~20220912A) using three different methods (PFA, LSP and H-test), but~yielded null results. For~the three sources with more than 1000 detected bursts (except for 20200120E), in-depth searches considering burst properties such as pulse width, peak flux, fluence, and~brightness temperature also fail to uncover any conspicuous~periodicity. 

These negative results for the four active repeaters are not isolated. Independent investigations of the same sources have likewise reported null findings. Li~et~al.~\cite{Li:2021hpl} conducted a systematic search for short-timescale periodicities (from 1 ms to 1000 s) in FRB~20121102A, using 1652 bursts observed with FAST, and~reported no significant detection. Nimmo~et~al.~\cite{Nimmo:2022dra} analyzed a burst storm from FRB 20200120E, and~found no strict periodicity in the burst arrival times, nor any evidence for periodicity in the source's activity between observations. Zhang~et~al.~\cite{Zhang:2023eui} conducted a short-term period search for FRB 20220912A and found no periodic signal in the range from 1 ms to 1000~s.

Among these four sources, FRB~20201124A has received particularly intensive scrutiny, yet no clear consensus on short-term periodicity has emerged. Niu~et~al.~\cite{Niu:2022fgq} examined more than 800 bursts from FRB~20201124A observed with FAST during an extremely active episode, and~found no strong evidence for a spin period in the 1 ms to 100 s range. Similarly, Xu~et~al.~\cite{Xu:2021qdn} analyzed 1863 bursts from this source, also detected by FAST, and~found that the burst arrival times exhibit substantial irregular short-term variations with no evidence of periodic modulation. Although~Du~et~al.~\cite{Du:2025lyg} reported a $\sim$1.7 s periodicity in the burst arrival times, this result has been questioned by Gazith \& Zackay~\cite{Gazith:2025gym}, who re-analyzed the same dataset and found no significant periodic signal. Furthermore, even if the $\sim$1.7 s periodicity were genuine, it would stand in marked contrast to the $\sim$26.24‑day periodicity observed in the Faraday RM variations of the same source~\cite{Xu:2025hal}.

\textls[-15]{Against this backdrop of largely null or controversial results in the millisecond‑to‑second} domain, the~recently discovered hyper‑active repeater FRB~20240114A offers a valuable new test bed. This source was monitored by FAST over 214 days from January to August 2024, yielding 11553 bursts, 3196 of which occurred within a single observing day (12 March 2024)~\cite{Zhang:2025qzn}. This dataset constitutes the largest burst sample from any FRB source to date, offering a unique opportunity for high-sensitivity periodicity searches. Katz~\cite{Katz:2025foo} performed a periodogram analysis on the bursts recorded during the 4.34 h window on the most active day, and~found no significant periodic signal, even when spin-down was~considered.

The absence of rotational periodicity has important implications. As~noted by Katz~\cite{Katz:2022sqt}, the~radio emission from any rotating magnetized object must be periodic at its rotational frequency unless the magnetic field is accurately dipolar and aligned with the rotation axis, which is an implausible scenario given the behavior of known pulsars. Rajwade \& Karastergiou~\cite{Rajwade:2025geh} proposed an alternative test based on polarization angle periodograms, which can detect an underlying rotation period even when the burst arrival times themselves appear aperiodic due to a large duty cycle. {Using this method, Liu~et~al.~\cite{Liu:2026zxx} searched for polarization-angle periodicity in three active FRB sources (FRB 20201124A, FRB 20220912A, and~FRB 20240114A) but found null results. Using long-term uGMRT polarization monitoring of FRB 20180916B, Bethapudi~et~al.~\cite{Bethapudi:2025iqb} found the polarization angle to be stable on timescales shorter than four hours but variable with activity phase and across cycles; they used these constraints to rule out fast-precession models for its progenitor. These early applications illustrate both the promise of PA-based periodicity searches and the need for larger, well-calibrated polarization datasets.}

Although periodicities have been successfully identified in some FRBs, several factors complicate the search for periodic signals in FRB~data:
\begin{itemize}
    \item\textbf{Sparsity of detections:} The number of detected bursts per source, even for active repeaters, is often insufficient for robust period detection over multiple cycles. While FRB~20240114A has produced over ten thousand bursts~\cite{Zhang:2025qzn}, most repeating FRBs have far fewer~detections.
    
    \item\textbf{Non-stationary activity:} The 1.7 s periodicity in FRB~20201124A was present only on two out of 49 days, indicating that periodic behavior may be transient and condition-dependent~\cite{Du:2025lyg}. Similarly, the~4.605-day candidate period in FRB~20121102A is only significant for higher-fluence bursts~\cite{Li:2024ckk}.
    
    \item\textbf{Multiple timescales:} FRB activity appears to exhibit periodicity on diverse timescales: milliseconds (intra-burst structure), seconds (burst arrival periodicity), days (activity cycles), and~months (long-term modulation). Distinguishing genuine periodic signals from aliases and sampling artifacts remains methodologically challenging.
\end{itemize}

\subsection{Theoretical Models for FRB~Periodicity}

Several theoretical frameworks have been advanced to explain the periodicity observed in repeating FRBs, encompassing a broad spectrum of physical origins, from~orbital dynamics and relativistic precession in binary systems, to~intrinsic spin or magnetospheric evolution of isolated neutron stars, and~even more exotic scenarios involving interactions with planetary debris or black hole–accretion disk systems. To~provide a coherent overview, we organize the proposed models below according to their primary driving mechanism and assess each against the available multi-wavelength and polarimetric~constraints.

\begin{itemize}

\item \textbf{Binary orbital modulation}: Ioka \& Zhang~\cite{Ioka:2020azq} pioneered the binary-comb model for FRB~20180916B, attributing its 16-day periodicity to the orbital motion of a highly magnetized pulsar interacting with a companion's wind, which periodically creates a transparent funnel for the radio emission. Wang~et~al.~\cite{Wang:2022mux} extended this framework to FRB~20201124A, showing that its dramatic RM variations, depolarization, and~circular polarization can be naturally reproduced by a magnetar orbiting a Be star with a decretion disk. Xu~et~al.~\cite{Xu:2025hal} provided strong evidence for the binary nature of FRB~20201124A by detecting a 26.24-day periodicity in its RM evolution over 14~consecutive cycles, robustly ruling out isolated precession scenarios in favor of orbital motion within a magnetized environment. Meanwhile, Lan~et~al.~\cite{Lan:2023vsu} systematically tested competing models for FRB~20180916B and found that while the ultralong rotation model best matches the stable period, it fails to reproduce the observed RM changes; they therefore proposed a hybrid picture combining a slowly rotating neutron star with a massive star companion, reconciling both the periodic activity and the RM evolution. Collectively, these studies point toward a growing consensus that many repeating FRBs reside in binary systems, where orbital dynamics and circumstellar plasma jointly shape the observed periodicities and polarization~properties.

\item \textbf{Precession in binary systems}: A second class of binary-related models invokes spin-axis precession of the neutron star, induced by the gravitational torque of a companion. Two specific variants have been explored in detail. The~orbit-induced spin precession model, advanced by Yang \& Zou~\cite{Yang:2020qxt}, suggests that the 16-day periodicity of FRB~20180916B arises naturally for plausible binary parameters, the~neutron star's spin axis precesses with a period of $10-20$ days, sweeping the emission beam across our line of sight only during specific precessional phases. While this model readily produces a narrow active window, it predicts short-lived tight binaries, which may conflict with population statistics. In~contrast, the~geodetic precession model attributes the modulation to general-relativistic frame-dragging effects, which cause the spin axis to precess with a period matching the observed $\sim$16-day signal. Wei~et~al.~\cite{Wei:2021vco} systematically compared these and other competing mechanisms—including free precession, radiation-driven precession, and~fall-back disk precession—against X-ray and gamma-ray constraints, and~concluded that geodetic precession is the most viable explanation for FRB~20180916B, whereas the others are each ruled out by at least one observational~constraint.

\item \textbf{Intrinsic models of isolated neutron stars}: Models that do not invoke a binary companion have also been proposed, although~they often face additional challenges in reproducing the full observational picture. For~FRB~20180916B, Lan~et~al.~\cite{Lan:2023vsu} considered the possibility that the periodic activity is driven by the ultra-long spin period of a slowly rotating neutron star. Among~the models they examined, this scenario appeared the most probable for explaining the periodicity alone, but~it cannot by itself account for the observed RM variations. This limitation motivated the hybrid binary+rotation proposal mentioned above. A~separate intrinsic channel involves a precessing magnetar, where a misalignment between the magnetic and rotational axes causes the radio beam to sweep across our line of sight with the precession period~\cite{Wei:2021vco}. For~FRB~20121102A, the~$\sim$160-day period could thus be interpreted as the precession period, while a candidate 4.6-day period might correspond to the magnetar's rotation. In~addition, Sob'yanin~\cite{Sobyanin:2020zqp} proposed that the 16.35-day and 157-day periodicities of FRB~20180916B and FRB~20121102A can be explained by forced precession of a neutron star, driven by an anomalous electromagnetic torque arising from non-corotational currents. This mechanism requires an internal magnetic field of $10^{14}$$\sim$$10^{15}$ G, characteristic of a typical~magnetar.

\item \textbf{Neutron star interaction with asteroid belt or planet}: Repetitive FRBs may also arise from gravitational interactions between a neutron star and an asteroid belt surrounding a massive companion~\cite{Deng:2024itt}. The~neutron star's tidal force perturbs the belt, causing asteroids to fall periodically onto the neutron star and produce radio bursts. Although~the active phase is finite because the belt is gradually depleted over successive passages, the~periodicity can persist for at least eight neutron star orbital periods. Alternatively, periodic FRBs could originate from a magnetized neutron star interacting with a planet on a highly eccentric orbit~\cite{Kurban:2021dmq}. At~each pericenter passage, the~planet is partially disrupted by tidal forces, and~the resulting debris accretes onto the neutron star, generating bursts via the Alfvén-wing mechanism. This model naturally accommodates periods ranging from days to hundreds of days, with~the precise value set by the orbital~parameters.

\item \textbf{Black hole–accretion disk system}: Katz~\cite{Katz:2024ecq} demonstrated that the dual periodicities of FRB~20121102A can be consistently interpreted within the framework of an intermediate-mass black hole with an accretion disk: the short-term 4.605-day period corresponds to the orbital motion of a star around the black hole, while the long-term 157-day period is identified as the precession period of the accretion disk~itself.

\end{itemize}

\subsection{Future~Directions for Periodicity Studies}

A critical frontier in FRB science is deciphering the physical origin of periodic activity, a~phenomenon that provides a rare window into the extreme environments around compact objects. Distinguishing among rotation, precession, and~binary orbital models is not merely a taxonomic exercise; it has direct implications for our understanding of neutron star magnetospheres, binary evolution, and~the potential link between FRBs and gravitational‑wave sources. Thus, the~ability to decisively discriminate among these competing scenarios using future observations stands as a central challenge for the field. Several promising approaches have been proposed:

\begin{itemize}
    \item\textbf{Polarization diagnostics:} The evolution of polarization angle over the activity cycle serves as a key discriminant between rotation, precession, and~orbital scenarios. In~a rotating magnetosphere, the~polarization angle is expected to track the rotational phase, whereas precession models produce a qualitatively distinct pattern~\cite{Rajwade:2025geh}.

    \item\textbf{Multi-frequency monitoring:} {Chromatic activity windows, exemplified by FRB 20180916B, can discriminate among models. The~observed narrower-at-higher-frequency windows disfavor simple wind-opacity scenarios, while binary/geometric models must be tested against the chromatic window widths~\cite{Gopinath:2023fty,Braga:2024ntx}.}

    \item\textbf{X-ray and \boldmath$\gamma$-ray follow-up:} While multiple models may yield analogous periodic signatures, multi‑wavelength observations at X‑ray and $\gamma$‑ray energies offer an independent diagnostic avenue to discriminate among them~\cite{Tavani:2020hqs,Wei:2021vco,Gouiffes:2024ouq}.

    \item\textbf{Gravitational wave searches:} Compact binary systems hosting neutron stars are potential sources for gravitational‑wave detectors such as LISA and LIGO. The~associated observation of FRBs and gravitational-waves  provides an independent route to testing the binary hypothesis~\cite{Yang:2020qxt,LIGOScientific:2022jpr,LIGOScientific:2024avz}.
\end{itemize}

\section{Polarization}\label{sec:polarization}

\subsection{Observational~Properties}

Polarization measurements serve as a fundamental diagnostic tool for deciphering the physical nature of FRBs. The~polarization state of FRB emission encodes crucial information about the magnetic field geometry in the emitting region, the~underlying particle acceleration processes, and~the properties of magnetized plasma along the line of sight. From~the Faraday RM and the evolution of the polarization angle, one can infer the magnetic field strength and orientation near the burst source, probe the electron density and magnetic field distributions in both the host galaxy and intervening media, and~even constrain the emission geometry and coherent radiation mechanisms. Furthermore, statistical analyses of large polarization samples offer unique opportunities to distinguish the physical origins of repeating and non‑repeating FRBs, to~unveil how diverse environments shape their polarization signatures, and~to use FRBs as cosmological probes of intergalactic and galactic magnetic fields. In~recognition of this potential, several major observational campaigns have devoted considerable resources to FRB polarization monitoring over recent years. Here, we synthesize the key findings from the latest large‑sample surveys, which have substantially advanced our understanding of FRB polarization properties. A~summary of the polarization characteristics for representative FRBs is provided in Table~\ref{tab:polarization}.

\begin{table}[H]
\caption{Polarization properties for representative~FRBs.}
\label{tab:polarization}
\begin{adjustwidth}{-\extralength}{0cm}

\begin{tabularx}{\linewidth}{L L l L L L}
\toprule
\textbf{FRB} & \textbf{Redshift} & \textbf{Linear Pol.} & \textbf{Circular Pol.} & \textbf{RM (rad m\boldmath$^{-2}$)} & \textbf{References} \\
\midrule
20121102A & $0.193$ & Frequency-dependent & Up to $\sim$64\%& $\sim$$1.5\times10^5$ & \cite{Michilli:2018zec,Hilmarsson:2020npc,Plavin:2022ost,Feng:2022ill,Feng:2022fap} \\
20180301A & $0.3304$ & $>$70\% (PA swing) & $<$3\% & $\sim$550 & \cite{Luo:2020kwy}\\
20180916B & $0.0337$ & $\gtrsim$80\% & $\lesssim$15\% & Variable & \cite{Nimmo:2020sva,Mckinven:2022fxn} \\
20201124A & ${0.09795}$ & Orthogonal modes & $\sim$90.9\% & Variable & \cite{Xu:2021qdn,Niu:2024mkg,Jiang:2024nnn} \\
20221022A & $65$ Mpc & $\sim$95\% (PA swing) & $<$10\% & $\sim$$-$40 & \cite{Mckinven:2024sbg} \\
20190520B & $0.241$ & Frequency-dependent & Low/Undetected & Sign reversals & \cite{Anna-Thomas:2022yvr,Feng:2022ill} \\
20220529A & $0.1839$ & Variable & Variable & Sudden change & \cite{Li:2025ckl}\\
20220912A & $0.0771$ & $>$90\% & Up to 58\% & Very stable & \cite{Zhang:2023eui,Feng:2023qym}\\
20230708A & $0.1050$ & $\sim$68\% (average) & Up to $\sim$75\% & $\sim$$-$6.9 & \cite{Dial:2024ljo} \\
20240114A & $0.1300$ & $>$90\% (PA swing) & Up to $\sim$65\% & Variable & \cite{Xie:2024fgo,Wang:2026ssk}\\
\bottomrule
\end{tabularx}
\end{adjustwidth}
\end{table}

\textbf{FRB 20121102A}: This is the first repeating FRB discovered and remains one of the most intensively studied sources. It is associated with a persistent radio source and resides in a low-metallicity, star-forming dwarf galaxy at $z=0.193$ \cite{Chatterjee:2017dqg,Tendulkar:2017vuq}. Michilli~et~al.~\cite{Michilli:2018zec} reported an extreme RM of $\sim$$1.5\times10^5\,$rad\,m$^{-2}$ for bursts from this source, indicating a highly magnetized plasma environment. Subsequent observations have revealed a strong frequency dependence of the burst polarization. The~linear polarization fraction rises steeply with frequency, increasing from $\sim$15\% at 1.7 GHz to $\sim$85\% at 3 GHz and reaching nearly $100\%$ at 5 GHz~\cite{Plavin:2022ost,Hilmarsson:2020npc}. Similar frequency-dependent polarization trend has been found in some other sources, such as repeating FRB 20190520B~\cite{Feng:2022ill} and non-repeating FRB 20230526A~\cite{Uttarkar:2025ydq}. Furthermore, Feng~et~al.~\cite{Feng:2022fap} detected $\sim$64\% circular polarization in one burst from FRB 20121102A, demonstrating that at least some repeating FRBs can exhibit high circular~polarization.

\textbf{FRB 20180301A}: This repeating source is localized to a star-forming galaxy at \mbox{z = 0.3304}~\cite{Bhandari:2021pvj} and provides a landmark result for understanding FRB emission mechanisms~\cite{Luo:2020kwy}. The~bursts from this source exhibit rich and diverse polarization angle swings: some show sweeping-up, sweeping-down, or~complex up-and-down patterns, while others display a constant polarization angle. The~measured RMs range from about $522$ to $564\ {\rm rad\,m^{-2}}$, with~evidence of variation on a timescale of one day. No significant circular polarization is detected ($<$3\%). The~diversity of polarization angle swings, combined with the high linear polarization fractions ($>$70\%), strongly favors a magnetospheric origin for the radio emission, such as from a neutron star, and~disfavors the synchrotron maser model in relativistic shocks, which predicts a constant polarization~angle.

\textbf{FRB 20180916B}: Nimmo~et~al.~\cite{Nimmo:2020sva} presented high-time-resolution ($\sim$1 $\upmu$s) polarimetric observations of four bursts from this repeating source at 1.7 GHz. All four bursts were highly linearly polarized ($\gtrsim$80\%), but~showed no significant circular polarization ($\lesssim$15\%). In~contrast to the polarization angle swings observed in FRB 20180301A, FRB 20180916B exhibited a constant polarization angle both during and between bursts. On~timescales shorter than $100\,\mu$s, however, subtle polarization angle variations of a few degrees were observed. These results are most naturally explained by a magnetospheric emission model, as~opposed to models in which the emission originates from a relativistic shock at larger distances. Detailed observations of FRB 20180916B with the CHIME telescope have revealed substantial RM evolution, while the DM remains nearly unchanged, suggesting large‐scale magneto‐ionic fluctuations in its local environment~\cite{Mckinven:2022fxn}. This is consistent with the view that repeating FRB progenitors may be associated with young stellar~populations.

\textbf{FRB 20201124A}: This active repeater exhibits some of the most extreme polarization properties known. Analyzing 1863 bursts from FAST, Xu~et~al.~\cite{Xu:2021qdn} detected circular polarization in over half of the sample, with~a maximum of $\sim$75\%. They also found oscillations in fractional linear/circular polarization and in polarization angle versus wavelength. The~RM showed significant variation during the first 36 days but remained stable afterwards. All these features point to a complex, dynamically evolving magnetized local environment. Niu~et~al.~\cite{Niu:2024mkg} reported the first detection of polarization angle orthogonal jumps, a~phenomenon previously only observed from radio pulsars. This finding provides strong evidence that FRB emission originates from the complex magnetosphere of a magnetar. Furthermore, Jiang~et~al.~\cite{Jiang:2024nnn} reported that this repeating source emits $\sim$90\% circularly polarized radio pulses, the~highest circular polarization fraction ever measured for an FRB. The~detection of such high-degree circular polarization, together with the orthogonal-mode structure of its linear polarization, places strong constraints on FRB emission mechanisms and suggests that both magnetospheric radiation and propagation effects are required to shape the observed~polarization.

\textbf{FRB 20221022A}: This non-repeating source is localized to a nearby host galaxy at a distance of $\sim$65 Mpc, offering a remarkable opportunity to study its polarization properties at close range. McKinven~et~al.~\cite{Mckinven:2024sbg} report a notable $\sim$$130^\circ$ rotation of the polarization angle over its $\sim$2.5 ms burst duration, resembling the characteristic S-shaped evolution seen in many pulsars and some radio magnetars. However, unlike pulsars, FRBs typically exhibit minimal variability in their polarization angle curves, and~even when marked evolution is present, the~curves often deviate from the canonical shape predicted by the rotating vector model. The~detection of a pulsar-like polarization angle swing from this nearby FRB provides important clues to the nature of the source, supporting the neutron star progenitor scenario and disfavoring models involving distant~shocks.

\textbf{FRB 20190520B}: Bursts from this repeating source exhibit extreme and highly variable Faraday rotation measures (RMs), including two sign reversals that clearly indicate a reversal in the line-of-sight magnetic field direction~\cite{Anna-Thomas:2022yvr}. Furthermore, the~bursts show significant depolarization at frequencies below $\sim$1--3~GHz~\cite{Anna-Thomas:2022yvr}, a~trend that has been confirmed by follow-up observations with the FAST and GBT telescopes~\cite{Feng:2022ill}. These observational features can be naturally interpreted as propagation through a turbulent, magnetized plasma screen, such as the stellar wind from a binary companion. These findings provide compelling evidence for a dense and dynamically evolving magnetoionic environment in the immediate vicinity of the FRB~source.

\textbf{FRB 20220529A}: This repeating source is distinguished by a sudden change and recovery in its local magnetic environment, as~revealed by an extreme Faraday RM flare~\cite{Li:2025ckl}. Long-term monitoring with FAST and Parkes over 2.2 years shows that during the first 17 months, the~bursts exhibited a median RM of $17~{\rm rad\,m^{-2}}$ with a scatter of $101~{\rm rad\,m^{-2}}$. Then, the RM abruptly jumped to $\sim$$1977~{\rm rad\,m^{-2}}$ in one day, and~monotonically declined to its baseline range within about two weeks. During~the decline, the~linear polarization fraction temporarily dropped from $\sim$80\% to $\sim$27\%, later recovering to normal values. This transient RM increase implies that a dense, magnetized plasma clump crossed the line of sight. The~most plausible interpretation is a coronal mass ejection from a companion star in a binary system, although~turbulence in a supernova remnant or pulsar wind nebula cannot be completely ruled out. The~estimated event rate of $\sim$0.45 flares per year suggests that such RM transients may be detectable in other repeating FRBs, offering a new probe of their local magneto-ionic~environments.

\textbf{FRB 20220912A}: In sharp contrast to FRB 20190520B and FRB 20220529A, the~repeater FRB 20220912A exhibits a remarkably stable electromagnetic environment~\cite{Feng:2023qym,Ravi:2023mnf}. Feng~et~al.~\cite{Feng:2023qym} detected 128 bursts with the GBT in 1.4 h observation, finding a nearly zero and stable rotation measure ($-0.4\pm0.3\,\mathrm{rad\,m^{-2}}$) and high linear polarization ($>$90\%), with~56\% of bright bursts showing significant circular polarization (up to $\sim$58\%). These results are further confirmed by FAST observations~\cite{Zhang:2023eui}. Both studies find that the large fraction of circularly polarized bursts is unlikely due to propagation effects and instead points to an intrinsic magnetospheric radiation mechanism, such as coherent curvature radiation or inverse Compton scattering. The~source's low and stable RM contrasts with other active repeaters, indicating that a highly magnetized environment is not a prerequisite for extreme repetition, and~the energy budget poses constraints on magnetar~models.

\textbf{FRB 20240114A}: The hyperactive repeater FRB 20240114A exhibits even richer polarization features. Xie~et~al.~\cite{Xie:2024fgo} report GBT observations at $720\text{--}920$~MHz, detecting 437~bursts with a peak rate of 264~h$^{-1}$; they find high linear polarization ($>$90\%) and circular polarization (up to 65\%), polarization angle swings, while the joint linear-versus-circular polarization distributions closely resemble those of the other two repeaters (FRB 20201124A and FRB 20220912A), suggesting a common radiation mechanism. Wang~et~al.~\cite{Wang:2026ssk} release a large FAST polarimetric catalog based on 16~months of monitoring, revealing a stable DM but a significant ($\sim$200~rad~m$^{-2}$) decline in RM over 200~days, with~persistently high linear polarization ($>$76\%) and low circular polarization, and~no evidence of frequency-dependent depolarization. Together, these results establish FRB\,20240114A as a dynamically evolving source with a stable emission mechanism and a variable magneto-ionic~environment.

\textbf{FRB 20230708A}: This non-repeating source displays complex temporal and polarimetric structure, featuring a suggestive quasi-periodicity of $7.267$ ms and high circular polarization (up to 75\%) in its first bright sub-burst~\cite{Dial:2024ljo}. The~linear polarization fraction is 68.4\% on average over the burst and reaches nearly 99\% at its peak, indicating an exceptionally high degree of linear polarization. The~polarization angle varies smoothly across components, yet the data firmly rule out generalized Faraday rotation from a relativistic plasma. The~interconnected polarization‑angle evolution and sub‑structure strongly point to a magnetar progenitor, while disfavoring a millisecond pulsar or compact binary merger~origin.

The PA swings observed in FRB 20180301A~\cite{Luo:2020kwy}, FRB 20221022A~\cite{Mckinven:2024sbg}, and~FRB 20240114A~\cite{Xie:2024fgo} place strong constraints on the magnetic environments of FRB sources. Liu~et~al.~\cite{Liu:2025nlo} analyze the PA evolution of 1727 bursts from three active repeating FRB sources (FRB 20180301A, FRB 20201124A and FRB 20220912A) observed with FAST. They find that most bursts have a nearly constant PA, while 46 bursts can be fitted by the rotating vector model (RVM). However, the~RVM-derived parameters are mutually inconsistent and many RVM-like profiles have large duty cycles, indicating that the PA variations are not driven by a stable magnetic rotator. The~authors conclude that the magnetosphere of the central engine is likely dynamically evolving and continuously distorted by the \mbox{FRB emitter.}

The CHIME/FRB project provides the largest sample of FRBs with polarization measurements to date. Pandhi~et~al.~\cite{Pandhi:2024cuo} report a polarimetric analysis of 128 non-repeating FRBs from the first CHIME/FRB catalog, of~which 89 exhibit $>$6$\sigma$ linearly polarized detections. After~subtracting the Milky Way contribution, the~derived RM amplitudes range from 0.5 to 1160~rad~m$^{-2}$, with~a median of 53.8~rad~m$^{-2}$. The~RMs of most non-repeating FRBs are consistent with Milky-Way-like host galaxies, while their linear polarization fractions span a wide range from $\leq$10\% to 100\%, with~a median value of 63\%. FRBs exhibit considerable diversity in their burst morphologies and polarization position angle profiles, with~the majority (57\%) displaying a single-component structure and a constant polarization angle. No significant population-wide frequency-dependent depolarization is found between $400\text{--}800$ MHz (CHIME band) or when compared with 1.4 GHz (DSA-110 band) data, indicating that the spread in linear polarization is likely intrinsic to the \mbox{emission~mechanism.}

Ng~et~al.~\cite{Ng:2024bwl} present polarization measurements for 75 bursts from 28 repeating FRBs in the first CHIME/FRB catalog. Their analysis yielded 41 new RM determinations for 20 repeaters and 22 updated measurements for 8 previously studied ones, with~RM values ranging from $-1044$ to $+1348$~rad~m$^{-2}$. They find that virtually all repeating FRBs exhibit temporal RM variations and identify two distinct categories based on the ratio of the RM standard deviation to the mean RM magnitude: sources with stable environments and those with dynamic magnetoionic surroundings. Notably, they observe a stochastic–secular–stochastic evolution in FRB~20180916B over five years, disfavoring binary orbital modulation as its cause, and~report RM sign changes for FRBs 20200929C and 20190303A. A~comparison between repeaters and non-repeaters shows marginal evidence for a dichotomy in the line-of-sight magnetic field strength distribution, consistent with earlier work~\cite{Pandhi:2024cuo}. 

Zhou~et~al.~\cite{Zhou:2026dza} have extended this polarimetric analysis by systematically measuring the RM-scatter parameter $\sigma_{\mathrm{RM}}$, which quantifies frequency-dependent depolarization, for~28~repeaters and 70 non-repeaters from the first CHIME/FRB catalog. For~repeaters, they found that about 70\% of sources showing $\sigma_{\mathrm{RM}}\gtrsim 1\ \mathrm{rad\,m^{-2}}$, suggesting that most repeaters reside in complex magneto-ionic environments; however, the~absence of values above $10\ \mathrm{rad\,m^{-2}}$ is likely a selection bias due to CHIME's limited bandwidth. For~non-repeaters, no significant depolarization ($\sigma_{\mathrm{RM}}\gtrsim 5\ \mathrm{rad\,m^{-2}}$) is detected; roughly half of the bursts are best described by a constant polarization fraction The authors caution that CHIME's narrow frequency coverage may obscure weaker depolarization effects, and~underscore that future ultra-wideband polarimetry will be pivotal for overcoming these~biases.

The Deep Synoptic Array (DSA-110) provides complementary polarization measurements at 1.4~GHz. Sherman~et~al.~\cite{Sherman:2024iev} reported a full-polarization analysis of 25 non-repeating FRBs. Faraday rotation measures (RMs) were detected for 20 FRBs, with~magnitudes ranging from 4 to 4670~rad~m$^{-2}$. Remarkably, 15 of the 25 FRBs are consistent with 100\% polarization, with~10 exhibiting high ($\geq$70\%) linear polarization fractions and two showing high ($\geq$30\%) circular polarization fractions. These results disfavor multipath RM scattering as a dominant depolarization mechanism, but~instead support a scenario in which FRB emission is intrinsically highly linearly polarized, while propagation effects can convert linear polarization into circular~polarization.

Uttarkar~et~al.~\cite{Uttarkar:2025ydq} conducted a depolarization census of 12 non-repeating FRBs detected by the ASKAP, and~found a strongly depolarized event (FRB 20230526A) showing a decrease in linear polarization fraction from $\sim$60\% at 1440 MHz to $\sim$20\% at 1110 MHz. This depolarization has been attributed to multipath propagation in a surrounding complex magneto-ionic environment. This finding, however, stands in contrast to the conclusion of Sherman~et~al.~\cite{Sherman:2024iev}, who argue that multipath RM scattering is not the dominant depolarization mechanism for the population, and~the observed spread in polarization fractions is largely intrinsic to the emission process~itself.

In summary, polarization observations of FRBs provide direct probes of magnetic field structures and emission physics. Most bursts are highly linearly polarized, with~fractions reaching up to $\sim$100\%, indicating a coherent radiation mechanism. The~Faraday rotation measure (RM) traces magnetized plasma along the line of sight; repeating FRBs frequently exhibit RM variability, including sign reversals and transient RM flares, which point to dynamically evolving local environments, although~some sources show remarkably stable RMs. Circular polarization is generally weak or absent, yet certain repeaters—such as FRB~20201124A—display extreme circular fractions up to $\sim$90\%. Moreover, polarization angle swings resembling pulsar-like behavior have been observed in sources like FRB~20180301A and FRB~20221022A. Large-sample surveys have further revealed that the RM distributions of non-repeaters and repeaters may differ, though~the relative roles of intrinsic emission mechanisms versus propagation effects (e.g., depolarization, mode conversion) remain actively~debated.

\subsection{Theoretical~Mechanisms}\label{sec:mechanisms}

Qu and Zhang~\cite{Qu:2023ibp} systematically explored a variety of polarization mechanisms within the framework of magnetar theoretical models. Based on the emission site, these models can be classified into two distinct categories: those operating inside the magnetosphere and those arising from regions outside~it. 

Inside the magnetosphere, two primary coherent emission mechanisms have been considered: coherent curvature radiation and inverse Compton scattering by charged bunches. Both mechanisms yield 100\% linear polarization for an on-axis viewing geometry, whereas off-axis lines of sight naturally introduce circular polarization. The~observed lack of circular polarization in the majority of bursts implies that the emitting bunches must possess sufficiently large transverse dimensions, such that most FRBs are viewed on-axis. Furthermore, resonant cyclotron absorption within the magnetosphere can also generate significant circular polarization if the electrons and positrons exhibit an asymmetric Lorentz factor~distribution.

Outside the magnetosphere, the~synchrotron maser mechanism operating at relativistic shocks typically yields emission that is highly linearly polarized. While circular polarization can arise at off-beam viewing angles, the~associated flux is severely suppressed, making such bursts undetectable over cosmological distances. Furthermore, synchrotron absorption in a nebula with an ordered magnetic field may further reduce the degree of circular polarization. In~contrast, cyclotron absorption in a strongly magnetized medium can produce significant circular~polarization.

Beyond these intrinsic emission mechanisms, propagation effects can also profoundly alter the observed polarization properties. In~a relativistic plasma, the~normal modes of wave propagation are not necessarily purely circular, leading to generalized Faraday rotation or Faraday conversion, where conversion between linear and circular polarizations occurs~\cite{Vedantham:2019wec,Cho:2020gtg}. This effect has been proposed as a possible origin for the circular polarization seen in some repeating FRB sources~\cite{Uttarkar:2025ydq}. In~a non‑relativistic cold plasma, by~contrast, the~natural modes are circularly polarized, so only standard Faraday rotation between the two orthogonal linear components takes place~\cite{Vedantham:2019wec}. Significant Faraday conversion may be realized in dense magnetized environments such as binary systems or supernova remnants~\cite{Qu:2023ibp}. Cyclotron absorption in a strongly magnetized medium can also generate substantial circular polarization~\cite{Qu:2023ibp}, but~the required physical conditions are stringent and this mechanism may not be universally applicable. Zhao and Wang~\cite{Zhao:2024srw} further proposed that circular polarization could arise from magnetospheric propagation effects induced by relativistic plasma; however, the~conditions needed to produce a high degree of circular polarization are rarely~satisfied.

In addition to these propagation-induced polarization changes, there are several distinct mechanisms that specifically act to reduce the observed linear polarization fraction, a~phenomenon often referred to as depolarization. These mechanisms are not mutually exclusive and can operate simultaneously depending on the source environment and geometry. They are summarized as follows:

\begin{itemize}
    \item \textbf{Multipath propagation in magnetized scattering screens.} The multipath propagation scenario attributes depolarization to the superposition of wave components traveling along distinct trajectories through a turbulent, magnetized scattering screen, where differential Faraday rotations accumulated along different lines of sight destroy the net linear polarization upon coherent summation~\cite{Beniamini:2021yto}. This framework is further developed by Yang~et~al.~\cite{Yang:2022wgs}, who establish a quantitative connection between the RM scatter, temporal scattering timescale, and~persistent radio emission. Uttarkar~et~al.~\cite{Uttarkar:2023cnk} provide observational support by demonstrating that spectral depolarization in both one-off and repeating FRBs can serve as a diagnostic for dense, turbulent, and~magnetized ionized plasma in the progenitor's vicinity. {Such scattering-induced depolarization and the flattening of polarization angle curves in scattered tails are also observed in Galactic pulsars~\cite{Jing:2025fxl}}

    \item \textbf{RM scattering in turbulent magneto-ionic media.} Closely related yet distinct, RM scattering specifically refers to frequency-dependent polarization angle fluctuations caused by spatial or temporal variations in the Faraday depth across the emitting region or along the sight line; such fluctuations produce a characteristic wavelength-dependent suppression of linear polarization that is particularly prominent in active repeating bursts such as FRB 20121102A. Feng~et~al.~\cite{Feng:2022ill} first identified a strong linear correlation between RM scattering parameter $\sigma_{\rm RM}$ and the scattering timescale $\tau_{\rm sca}$ in active repeaters, and~Yang~et~al.~\cite{Yang:2022wgs} further confirm that both quantities originate from the same inhomogeneous magnetized plasma region, reinforcing RM scattering as a robust depolarization~channel.

    \item \textbf{Faraday conversion between polarization modes.} Faraday conversion offers a fundamentally different mechanism, in~which linear and circular polarization modes are mutually converted when the FRB signal traverses a strongly magnetized and dense plasma environment, such as the magnetosphere of a companion star in a binary system. Xia~et~al.~\cite{Xia:2023jkg} show that under extreme plasma column densities, this interconversion can lead to rapid frequency oscillations in circular polarization and, consequently, a~severe reduction in the observed polarization degree, providing a distinctive observational signature that is not explained by Faraday rotation alone. Furthermore, a~sufficiently significant RM reversal can be produced at large magnetic inclinations and the RM variation is very diverse, which can well explain the RM variations observed in e.g.,~FRB 20180916B~\cite{Mckinven:2022fxn} and FRB 20201124A~\cite{Xu:2021qdn}.
\end{itemize}

\subsection{Implications for Progenitor~Models}\label{sec:implications}

The accumulated polarimetric evidence strongly favors a magnetar origin for at least a substantial fraction of FRBs, and~magnetars remain the most favored progenitor class among the many proposed models~\cite{Popov:2023bpk}. The~diversity of polarization signatures, particularly in repeating FRBs, strongly favors a magnetospheric origin, wherein the emission is generated within a dynamic, highly magnetized plasma~\cite{Wang:2025tmy}. For~instance, the~diverse polarization angle swings observed in FRB~20180301A~\cite{Luo:2020kwy}, the~nearly constant polarization angle with subtle variations in FRB~20180916B~\cite{Nimmo:2020sva}, and~the pulsar-like polarization angle swings in FRB~20221022A~\cite{Mckinven:2024sbg} all point toward magnetospheric emission. These distinct behaviors likely reflect different viewing geometries, emission altitudes, or~local plasma conditions within the magnetosphere, further strengthening the case for a magnetospheric~origin.

FRB polarization properties exhibit both intriguing similarities to and notable differences from those of Galactic pulsars. Sherman~et~al.~\cite{Sherman:2024iev} compared the polarization properties of FRB subpopulations with those of Galactic pulsars, finding that although FRB polarization fractions are typically higher and span a wider range than pulsar single pulses, they closely resemble those of the youngest pulsars (characteristic ages $<10^5$\,yr). The~detection of orthogonal jumps in the polarization angle of FRB~20201124A, a~phenomenon previously observed only in radio pulsars, further solidifies the connection between FRBs and neutron-star magnetospheres~\cite{Niu:2024mkg}. Such polarization angle jumps are naturally interpreted as transitions between orthogonal polarization modes in a magnetospheric plasma, providing direct evidence that the underlying emission physics may be shared between FRBs and young pulsars, despite the dramatically higher energy scales of~FRBs.

One of the key questions in FRB astrophysics is whether repeating and apparently non-repeating FRBs share a common physical origin, and~polarization provides important constraints on this issue. Pandhi~et~al.~\cite{Pandhi:2024cuo} found marginal evidence that non-repeating FRBs exhibit tighter lower limits on the host's electron-density-weighted line-of-sight magnetic field strength compared to repeating FRBs. In~contrast, Uttarkar~et~al.~\cite{Uttarkar:2025ydq} argued that their analysis supports a scenario in which repeaters and non-repeaters share a common origin, with~non-repeaters representing an older population relative to repeating sources. However, the~polarization distributions of repeating and non-repeating FRBs may still differ, suggesting either distinct emission mechanisms or varying levels of depolarization in the local environments~\cite{Niu:2024mkg}. Reconciling these results likely requires larger samples of bursts with high-quality polarimetry to disentangle genuine evolutionary trends from observational biases and intrinsic~diversity.

\subsection{Future~Directions for Polarimetry}

Several key avenues for advancing FRB polarimetry are readily apparent. First, enlarging the sample of FRBs with robust polarization measurements is imperative for establishing population-level statistics and identifying potential subclasses. Ongoing surveys such as CHIME/FRB and DSA-110 continue to expand the observational catalog, while next-generation facilities like the Square Kilometer Array (SKA) will deliver transformative sensitivity and bandwidth.
Second, high-time-resolution polarimetry, as~demonstrated by Nimmo~et~al.~\cite{Nimmo:2020sva} for FRB 20180916B, can resolve micro-structural features and rapid variations in polarization angle (PA), offering stringent constraints on emission models. Observations achieving microsecond (or better) temporal resolution will prove especially instrumental in this regard. Third, polarimetric observations spanning a broad radio-frequency range are critical for disentangling intrinsic polarization from propagation-induced effects. The~frequency-dependent behavior of the polarization fraction, PA, and~Faraday rotation measure (RM) encodes valuable information about both the underlying emission mechanism and the surrounding magneto-ionic environment. Fourth, coordinated multi-wavelength campaigns, encompassing X-ray, optical, and~gamma-ray regimes, can provide complementary constraints on the progenitor and its immediate vicinity. The~detection of an FRB-like burst from the Galactic magnetar SGR J1935+2154~\cite{Bochenek:2020zxn} has firmly established the first direct link between FRBs and magnetars; future associations of this kind will be invaluable. Finally, theoretical modeling must keep pace with the accelerating observational progress. Detailed treatments of magnetospheric polarization, incorporating geometric effects, propagation, and~absorption, will be essential for interpreting the increasingly rich and complex datasets.

\section{Conclusions and Future~Prospects}\label{sec:summary}

In this review, we have synthesized the current state of statistical knowledge on FRBs, drawing from the rapidly growing observational samples provided by radio telescopes worldwide, such as CHIME, ASKAP, DSA-110, FAST, and~other facilities. The~statistical properties of FRBs has matured considerably over the past few years, transitioning from small-sample exploratory studies to large, homogeneous analyses that yield robust empirical constraints. Nevertheless, the~physical interpretation of these statistical patterns remains a subject of active debate, and~many fundamental questions remain~open.

For non-repeating FRBs, the~empirical landscape is now well defined. The~host-galaxy dispersion measure $\mathrm{DM}_{\mathrm{host}}$ follows a lognormal distribution with a median of $\sim$150~$\mathrm{pc\,cm^{-3}}$, consistent with cosmological simulations, although~its correlation with host properties remains contentious. The~energy function is universally described by power law with a high-energy cutoff, with~index $\alpha \simeq 1.8\text{--}1.9$ and cutoff energy $\log(E_{\mathrm{max}}/\mathrm{erg}) \simeq 42$, a~result that is remarkably stable across samples and methods. {The redshift evolution, however, is less settled. Many CHIME-based analyses disfavor a pure SFH model and prefer suppressed evolution or Gyr-scale delays, but~other treatments that incorporate injection-based selection functions find little or no delay or no significant constraint on the evolution relative to the SFH. Thus, the evidence for an older, delayed progenitor population is suggestive but assumption-dependent; young core-collapse magnetars could not be decisively excluded as a dominant population by current statistics.} Deep follow-up campaigns have placed stringent upper limits on the repetition rates of apparently non-repeating FRBs, typically $\lesssim$$10^{-2}~\mathrm{h^{-1}}$, yet population models suggest that the intrinsic repeater fraction may be much higher, possibly approaching unity, implying that the apparent dichotomy between repeaters and non-repeaters is largely an observational selection~effect.

Repeating FRBs, by~contrast, exhibit a far richer and more complex statistical behavior. Their energy, fluence, and~flux-density distributions routinely deviate from simple power laws, requiring bent power-law, thresholded power-law, broken power-law, or~bimodal lognormal models to capture turnovers, breaks, and~multi-component structures. Waiting-time distributions are usually multimodal, typically described by Weibull or multi-peaked lognormal functions, reflecting burst clustering on multiple timescales from millisecond-scale substructure to hour-long activity cycles. Time-series analyses have uncovered persistent memory effects, scale-invariant fluctuations that follow the Tsallis $q$-Gaussian distribution, and~stochastic signatures that bear a striking resemblance to earthquakes and magnetar bursts. These dynamical fingerprints imply self-organized criticality in neutron-star crusts as the underlying trigger~mechanism.

Periodicity studies have revealed a growing number of sources with periodic activity on diverse timescales, ranging from milliseconds (intra-burst substructure), seconds (burst-arrival periodicities), days (activity cycles), to~months (long-term modulation). FRB 20180916B's 16.35-day cycle with frequency-dependent active windows, FRB 20121102A's $\sim$159-day and candidate 4.6-day periodicities, FRB 20201124A's 26.24-day RM periodicity and transient $\sim$1.7 s signal, and~especially the multi-timescale periodicities in FRB 20240114A collectively point toward a variety of physical mechanisms, such as binary orbital motion, spin precession, and~possibly intrinsic rotation. However, comprehensive searches for millisecond-to-second rotational periods in most active repeaters have yielded null results, posing a significant challenge to simple magnetar-rotation~models.

Polarization observations have established that FRB emission is intrinsically highly linearly polarised, with~fractions often exceeding $90\%$, indicative of coherent radiation mechanisms. Faraday rotation measures (RMs) probe the magneto-ionic environments, revealing striking diversity: some repeaters exhibit extreme RM variability, sign reversals, and~transient RM flares (e.g., FRB 20190520B, FRB 20220529A), while others show remarkably stable environments (e.g., FRB 20220912A). Circular polarization is generally weak but can reach extreme values in certain repeaters, such as FRB 20201124A ($>$90\%). Polarization angle swings resembling pulsar-like behavior have been detected in sources like FRB 20180301A and FRB 20221022A, strongly favoring magnetospheric emission over shock-based models. Large-sample comparisons reveal marginal differences in the RM distributions of repeaters and non-repeaters, though~the relative contributions of intrinsic emission physics versus propagation effects remain to be fully~disentangled.

\textls[-15]{Despite these substantial advances, several outstanding questions and \mbox{challenges remain:}}

\begin{itemize}
\item \textbf{The nature of FRB progenitors.} While magnetars are the leading candidate, the~diversity of host environments, ranging from dwarf star-forming galaxies to massive quiescent ellipticals and globular clusters, strongly suggests multiple formation channels. The~inferred time delay of $3$--$5$ Gyr relative to SFH points toward old progenitors, yet the detection of FRB-like bursts from the Galactic magnetar SGR 1935+2154 demonstrates that young magnetars can also produce such emission. Resolving this apparent tension requires larger samples of well-localized FRBs with robust host-galaxy characterization, as~well as direct multi-wavelength counterparts. {Wide-field surveys, such as CHIME and SKA, will play a important role in assembling the large, well-localized FRB samples needed to test progenitor scenarios.}

\item \textbf{The origin of periodic activity.} The physical driver of periodicity remains unclear. Distinguishing between binary orbital motion, neutron-star precession, and~intrinsic rotation demands multi-frequency, multi-epoch monitoring combined with polarimetric and RM tracking. The~chromatic behavior of FRB 20180916B strongly {strongly disfavors simple stellar-wind free–free absorption as the direct cause of the periodicity}, while the stable RM of FRB 20220912A and the RM periodicity of FRB 20201124A favor different scenarios. Definitive discrimination may require coordinated observations with next-generation facilities and, potentially, gravitational-wave searches for binary~companions.

\item \textbf{The emission mechanism.} The high linear polarization, complex polarization angle swings, and~extreme circular polarization in some repeaters provide strong constraints on coherent emission models. The~balance between intrinsic magnetospheric radiation (curvature radiation, inverse Compton scattering) and propagation-induced effects (Faraday conversion, cyclotron absorption, multipath scattering) remains poorly quantified. Ultra-wideband polarimetry and high-time-resolution observations are essential to disentangle these effects and to determine whether a single universal mechanism can explain the full diversity of polarization~signatures.

\item \textbf{The repeater--non-repeater dichotomy.} Whether all FRBs ultimately repeat at extremely low rates or whether a genuinely non-repeating population exists remains unresolved. While population models suggest that the intrinsic repeater fraction may exceed $50\%$ or even approach unity, temporal evolution analyses of the observed repeater fraction allow for the possibility of true one-off events. Continuous monitoring with wide-field telescopes and long-baseline campaigns are needed to settle \mbox{this question.}
\end{itemize}

Looking ahead, several key observational and methodological developments will shape the future of FRB statistics:

\begin{itemize}
\item \textls[-15]{\textbf{Wide-field surveys.} {The operating and forthcoming wide-field surveys, such as SKA~\cite{SKA:2018ckk}, DSA-2000~\cite{Hallinan:2023}, BURSTT~\cite{Lin:2022wgp}, STARE2~\cite{Bochenek:2020qry}, CASATTA~\cite{Gutierrez:2025}, CASM~\cite{Sanghavi:2026}, CHARTS~\cite{Cassanelli:2025} and BINGO~\cite{Abdalla:2021nyj,dosSantos:2023ipw}, will deliver FRB samples orders of magnitude larger than current ones, with~unprecedented sensitivity, bandwidth, and~localization precision.} These will enable robust population-level analyses, detailed studies of rare subclasses, and~statistically meaningful comparisons between repeaters \mbox{and~non-repeaters.}}

\item \textbf{High-time-resolution and ultra-wideband polarimetry.} Microsecond-resolution polarimetry and observations spanning from $\sim$100 MHz to several GHz will be critical for disentangling intrinsic emission from propagation effects, for~characterizing the frequency dependence of polarization and RM, and~for probing the microphysical conditions in the emission~region.

\item \textbf{Multi-wavelength and multi-messenger synergies.} Coordinated observations in X-ray, optical, and~gamma-ray bands, as~well as searches for gravitational-wave counterparts, will provide complementary constraints on progenitor models and environment properties. The~detection of more FRB-like bursts from Galactic magnetars or from nearby external galaxies would be a~game-changer.

\item \textbf{Advanced statistical methods.} The application of machine learning, Bayesian hierarchical modeling, and~non-parametric inference will be essential for handling the complex selection functions, multi-dimensional parameter spaces, and~correlated uncertainties inherent in FRB data. These methods will also enable more rigorous model comparison and parameter estimation, moving beyond simple goodness-of-fit~tests.

\item \textbf{Theoretical synthesis.} The statistical patterns reviewed here---from power-law energy functions to waiting-time multimodality, from~scale-invariant fluctuations to periodic modulations---demand a unified theoretical framework that can simultaneously account for the full diversity of observed behaviors. Self-organized criticality in magnetar crusts, coupled with magnetospheric emission and binary interactions, appears to be a promising avenue, but~detailed quantitative models that can be directly confronted with data are still lacking.
\end{itemize}

In conclusion, the~statistical study of FRBs has matured into a rich and multifaceted discipline that provides essential constraints on progenitor models, emission physics, and~cosmic environments. The~empirical foundations are now solid: we know that FRB energy distributions are universally power-law with a high-energy cutoff, that the population does not trace the star formation history, that repeating bursts exhibit complex temporal clustering and memory effects, and~that polarization properties point toward magnetospheric emission from neutron stars. Yet, the physical interpretations of these findings remain fluid, and~the field is poised for transformative advances with the next-generation observational facilities. The~ultimate goal, namely identifying the definitive progenitor and emission mechanism of FRBs, remains within reach, and~statistical analyses will continue to play a central role in this pursuit.

\vspace{+6pt}
\authorcontributions{Conceptualization and outline, H.-N.L.; literature collection, Y.S.; writing—original draft preparation, H.-N.L. (Sections \ref{sec:nonrepeater} and \ref{sec:periodicity}--\ref{sec:summary}) and Y.S. (Sections \ref{sec:introduction} and \ref{sec:repeater}); writing—review and editing, H.-N.L. and Y.S. All authors have read and agreed to the published version of the~manuscript.}

\funding{This work has been supported by the National Natural Science Fund of China under grant Nos. 12675062, 12347101 and 12275034; and the Natural Science Fund of Chongqing under grant no. CSTB2022NSCQ-MSX0357; and Yangzhou Science and Technology Planning Project in Jiangsu Province of China under grant no.~YZ2025233.}

\dataavailability{Data sharing is not applicable to this article as no new data were created or analyzed in this study.}

\conflictsofinterest{The authors declare no conflicts of~interest.}

\begin{adjustwidth}{-\extralength}{0cm}

\reftitle{References}

\PublishersNote{}
\end{adjustwidth}

\end{document}